\documentclass[11pt,a4paper,twosided]{article}

\usepackage{geometry}
\usepackage[utf8]{inputenc}
\usepackage[T1]{fontenc}

\usepackage[english]{babel}

\usepackage{microtype}

\usepackage{amsmath,amsthm,amssymb,stmaryrd,thmtools,upgreek}

\newtheorem{theorem}{Theorem}
\newtheorem{lemma}[theorem]{Lemma}

\theoremstyle{definition}
\newtheorem{definition}[theorem]{Definition}

\usepackage{graphicx}
\usepackage[obeyclassoptions,mode=tex]{standalone}
\usepackage{tikz}
\usetikzlibrary{backgrounds}
\usetikzlibrary{shapes.geometric}
\usetikzlibrary{positioning}
\usetikzlibrary{automata}
\usetikzlibrary{tikzmark}
\usetikzlibrary{patterns}
\usetikzlibrary{arrows}
\tikzset{every state/.style={minimum size=1pt}}
\usepackage{tikz-cd}

\usepackage{hyperref}
\usepackage[capitalise,noabbrev,nameinlink]{cleveref}
\usepackage[electronic,hyperref,xcolor,cleveref]{knowledge}
\knowledgeconfigure{notion}

\usepackage{booktabs}
\usepackage{varwidth}

\usepackage{xparse}
\usepackage{xpatch}
\usepackage{tokcycle}
\usepackage{ifthen}

\usepackage{bussproofs}

\usepackage{ensps-colorscheme}

\usepackage{fancyvrb}
\usepackage{enumitem}
\usepackage{wrapfig}

\usepackage{pifont}%
\newcommand{\cmark}{\ding{51}}%
\newcommand{\xmark}{\ding{55}}%
\newcommand{\OK}{\textcolor{A5}{\cmark}}
\newcommand{\KO}{\textcolor{C3}{\xmark}}
\newcommand{\UK}{\textcolor{C4}{?}}

\DefineVerbatimEnvironment{Highlighting}{Verbatim}{fontsize=\small,commandchars=\\\{\},frame=single,framesep=4mm,numbersep=1mm}
\newenvironment{Shaded}{}{}

\newcommand{\BuiltInTok}[1]{\textcolor{A5}{#1}}

\newcommand{\CommentTok}[1]{\textcolor{C1}{\textit{#1}}}

\newcommand{\ControlFlowTok}[1]{\textcolor{A5}{\textbf{#1}}}
\newcommand{\DataTypeTok}[1]{\textcolor{B2}{#1}}

\newcommand{\KeywordTok}[1]{\textcolor{A5}{\textbf{#1}}}
\newcommand{\NormalTok}[1]{#1}
\newcommand{\OperatorTok}[1]{\textcolor{A3}{#1}}

\newcommand{\StringTok}[1]{\textcolor{D2}{#1}}
\newcommand{\VariableTok}[1]{\textcolor{B4}{#1}}

\makeatletter
\newcommand\mathgr[1]{\tokcycle
  {\addcytoks{##1}}
  {\processtoks{##1}}
  {\ifcsname up\expandafter\@gobble\string##1\endcsname
   \addcytoks[1]{\csname up\expandafter\@gobble\string##1\endcsname}%
    \else\addcytoks{##1}\fi}
  {\addcytoks{##1}}{#1}%
  \expandafter\mathrm\expandafter{\the\cytoks}%
}
\makeatother

\NewDocumentEnvironment{proofof}{ m O{appendix} }{
    \ifcsname #1\endcsname
        \def\isInsideRestatedTheorem{1}
        \csname #1\endcsname*
    \fi
    \begin{proof}[Proof of {\cref{#1}} as stated on page {\pageref{#1}}]
        \phantomsection
        \label{#1:proof}
}{
        \ifthenelse{\equal{#2}{appendix}}{
        \marginpar{\vspace{-2em}\texttt{\small{\hyperref[#1]{$\triangleright$ Back to p.\pageref{#1}}}}}
        }{}
    \end{proof}
}

\NewDocumentCommand{\proofref}{ m }{
    \IfRefUndefinedExpandable{#1:proof}{}{
        \ifdefined\isInsideRestatedTheorem
        \else
            \marginpar{\vspace{0.6em}\texttt{\small{\hyperref[#1:proof]{$\triangleright$ Proven p.\pageref{#1:proof}}}}}
        \fi
    }
}

\NewDocumentCommand{\NewDocumentOrdering}{ m m m }{
    \expandafter\newcommand\csname #1leq\endcsname{
        \mathrel{\kl[#1]{#2}}
    }
    \expandafter\newcommand\csname #1lt\endcsname{
        \mathrel{\kl[#1]{#3}}
    }
    \knowledge{#1}{notion}
}

\NewDocumentCommand{\set}{ m }{\{ #1 \}}

\NewDocumentCommand{\Nat}{ }{\mathbb{N}}

\NewDocumentCommand{\seqof}{ m O{n \in \Nat} }{\left( #1 \right)_{#2}}

\NewDocumentCommand{\range}{ O{1} m }{[#1, #2]}

\NewDocumentCommand{\upset}{ O{} m }{{\uparrow_{#1} #2}}
\NewDocumentCommand{\dwset}{ O{} m }{{\downarrow_{#1} #2}}

\NewDocumentCommand{\factorial}{ O{} m }{
    \if\relax\detokenize{#1}\relax
        #2!
    \else
        (#2)!
    \fi
}

\newcommand{\circled}[2]{%
\hypertarget{#2}{}%
\tikz[baseline=(char.base),color=A2,thick]{%
\node[shape=circle,draw,inner sep=1pt,font=\tiny] (char) {#1};%
}}
\newcommand{\circleref}[2]{%
\hyperlink{#2}{%
\tikz[baseline=(char.base),color=A2,thick]{%
\node[shape=circle,draw,inner sep=1pt,font=\tiny] (char) {#1};}%
}}

\NewDocumentCommand{\FO}{ o }{\IfNoValueTF{#1}{\kl[\FO]{\mathsf{FO}}}{\kl[\FO]{\mathsf{FO}[{#1}]}}}
\knowledge{\FO}{notion}

\NewDocumentCommand{\MSO}{ o }{\IfNoValueTF{#1}{\kl[\MSO]{\mathsf{MSO}}}{\kl[\MSO]{\mathsf{MSO}[{#1}]}}}
\knowledge{\MSO}{notion}

\NewDocumentCommand{\hoaretriple}{ m m m }{\withkl{\kl[\hoaretriple]}{\{ #1 \} \; #2 \; \{ #3 \}}}
\knowledge{\hoaretriple}{notion}

\NewDocumentCommand{\NestedWords}{O{}}{\kl[\NestedWords]{\mathcal{W}_{#1}}}
\knowledge{\NestedWords}{notion}

\NewDocumentCommand{\PVars}{}{\kl[\PVars]{\mathbb{V}_{\text{pos}}}}
\knowledge{\PVars}{notion}

\NewDocumentCommand{\FunVars}{}{\kl[\FunVars]{\mathbb{V}_{\text{fun}}}}
\knowledge{\FunVars}{notion}

\NewDocumentCommand{\BVars}{}{\kl[\BVars]{\mathbb{V}_{\text{bool}}}}
\knowledge{\BVars}{notion}

\NewDocumentCommand{\OVars}{}{\kl[\OVars]{\mathbb{V}_{\text{list}}}}
\knowledge{\OVars}{notion}

\NewDocumentCommand{\Bools}{}{\kl[\Bools]{\mathbb{B}}}
\knowledge{\Bools}{notion}

\NewDocumentCommand{\OutputType}{O{\Sigma}}{\kl[\OutputType]{\mathcal{C}_{#1}}}
\knowledge{\OutputType}{notion}

\NewDocumentCommand{\hlprogram}{}{\kl[\hlprogram]{\mathsf{Program}}}
\knowledge{\hlprogram}{notion}

\NewDocumentCommand{\hlfun}{}{\kl[\hlfun]{\mathsf{Fun}}}
\knowledge{\hlfun}{notion}

\NewDocumentCommand{\hlstmt}{}{\kl[\hlstmt]{\mathsf{Stmt}}}
\knowledge{\hlstmt}{notion}

\NewDocumentCommand{\bexpr}{}{\kl[\bexpr]{\mathsf{BExpr}}}
\knowledge{\bexpr}{notion}

\NewDocumentCommand{\oexpr}{}{\kl[\oexpr]{\mathsf{LExpr}}}
\knowledge{\oexpr}{notion}

\NewDocumentCommand{\cexpr}{}{\kl[\cexpr]{\mathsf{CExpr}}}
\knowledge{\cexpr}{notion}

\NewDocumentCommand{\aexpr}{}{\kl[\aexpr]{\mathsf{AExpr}}}
\knowledge{\aexpr}{notion}

\NewDocumentCommand{\bBinOp}{}{\kl[\bBinOp]{\mathsf{BBin}}}
\knowledge{\bBinOp}{notion}

\NewDocumentCommand{\pCmpOp}{}{\kl[\pCmpOp]{\mathsf{PComp}}}
\knowledge{\pCmpOp}{notion}

\NewDocumentCommand{\pyield}{ m m }{\kl[\pyield]{\mathsf{pos}^{#1}_{#2}}}
\knowledge{\pyield}{notion}

\NewDocumentCommand{\hlif}{m m m}{\withkl{\kl[\hlif]}{\cmdkl{\mathsf{if}} \; #1 \; \cmdkl{\mathsf{then}} \; #2 \; \cmdkl{\mathsf{else}} \; #3}}
\NewDocumentCommand{\hlifnoelse}{m m}{\withkl{\kl[\hlif]}{\cmdkl{\mathsf{if}} \; #1 \; \cmdkl{\mathsf{then}} \; #2}}
\knowledge{\hlif}{notion}

\NewDocumentCommand{\hlyield}{m}{\withkl{\kl[\hlyield]}{\cmdkl{\mathsf{yield}} \; #1}}
\knowledge{\hlyield}{notion}

\NewDocumentCommand{\hlskip}{m}{\mathsf{skip}}

\NewDocumentCommand{\hlreturn}{m}{\withkl{\kl[\hlreturn]}{\cmdkl{\mathsf{return}} \; #1}}
\knowledge{\hlreturn}{notion}

\NewDocumentCommand{\hlletoutput}{m m m}{\withkl{\kl[\hlletoutput]}{\mathsf{let} \; #1 = #2 \; \mathsf{in} \; #3}}
\knowledge{\hlletoutput}{notion}

\NewDocumentCommand{\hlletboolean}{m m}{\withkl{\kl[\hlletboolean]}{\cmdkl{\mathsf{let~mut}} \; #1\cmdkl{~=~\bfalse \; \mathsf{in}} \; #2}}
\knowledge{\hlletboolean}{notion}

\NewDocumentCommand{\hlsettrue}{m}{\withkl{\kl[\hlsettrue]}{\; #1 \; \cmdkl{\leftarrow} \; \cmdkl{\mathsf{true}}}}
\knowledge{\hlsettrue}{notion}

\NewDocumentCommand{\hlfor}{m m m}{\withkl{\kl[\hlfor]}{\cmdkl{\mathsf{for}^{\rightarrow}} \; #1 \; \cmdkl{\mathsf{in}} \; #2 \; \cmdkl{\mathsf{do}} \; #3}}
\knowledge{\hlfor}{notion}

\NewDocumentCommand{\hlforRev}{m m m}{\withkl{\kl[\hlforRev]}{\cmdkl{\mathsf{for}^{\leftarrow}} \; #1 \; \cmdkl{\mathsf{in}} \; #2 \; \cmdkl{\mathsf{do}} \; #3}}
\knowledge{\hlforRev}{notion}

\NewDocumentCommand{\hlseq}{m m}{\withkl{\kl[\hlseq]}{#1 \; \cmdkl{;} \; #2}}
\knowledge{\hlseq}{notion}

\NewDocumentCommand{\btrue}{}{\kl[\btrue]{\mathsf{true}}}
\knowledge{\btrue}{notion}

\NewDocumentCommand{\bfalse}{}{\kl[\bfalse]{\mathsf{false}}}
\knowledge{\bfalse}{notion}

\NewDocumentCommand{\bnot}{m}{\withkl{\kl[\bnot]}{\cmdkl{\neg} #1}}
\knowledge{\bnot}{notion}

\NewDocumentCommand{\bbin}{ m m m }{\withkl{\kl[\bbin]}{#1 \; #2 \; #3}}
\knowledge{\bbin}{notion}

\NewDocumentCommand{\bcomp}{m m m}{\withkl{\kl[\bcomp]}{#1 \; #2 \; #3}}
\knowledge{\bcomp}{notion}

\NewDocumentCommand{\bapp}{m m}{\withkl{\kl[\bapp]}{\mathop{#1}(#2)}}
\knowledge{\bapp}{notion}

\NewDocumentCommand{\bliteq}{m m}{\withkl{\kl[\bliteq]}{#1 \; \cmdkl{=_{\text{lit}}} \; #2}}
\knowledge{\bliteq}{notion}

\NewDocumentCommand{\bgen}{m}{\withkl{\kl[\bgen]}{\mathopen{\cmdkl{\langle}} #1 \mathclose{\cmdkl{\rangle_{\text{b}}}}}}
\knowledge{\bgen}{notion}

\NewDocumentCommand{\ogen}{m}{\withkl{\kl[\ogen]}{\mathopen{\cmdkl{\langle}} #1 \mathclose{\cmdkl{\rangle_{\text{l}}}}}}
\knowledge{\ogen}{notion}

\NewDocumentCommand{\cchar}{m}{\withkl{\kl[\cchar]}{\mathsf{char}(#1)}}
\knowledge{\cchar}{notion}

\NewDocumentCommand{\clist}{m}{\withkl{\kl[\clist]}{\mathsf{list}(#1)}}
\knowledge{\clist}{notion}

\NewDocumentCommand{\olist}{m}{\mathop{\kl[\olist]{\mathsf{list}}(#1)}}
\knowledge{\olist}{notion}

\NewDocumentCommand{\hlfundef}{m m m}{\withkl{\kl[\hlfundef]}{\mathsf{def} \; #1(#2) \; \{ #3 \}}}
\knowledge{\hlfundef}{notion}

\NewDocumentCommand{\TPos}{O{} }{\kl[\TPos]{\mathsf{Pos}_{#1}}}
\knowledge{\TPos}{notion}

\NewDocumentCommand{\TBool}{}{\kl[\TBool]{\mathsf{Bool}}}
\knowledge{\TBool}{notion}

\NewDocumentCommand{\TUnit}{}{\kl[\TUnit]{\mathsf{Any}}}
\knowledge{\TUnit}{notion}

\NewDocumentCommand{\TOut}{O{}}{\kl[\TOut]{\mathsf{List}_{#1}}}
\knowledge{\TOut}{notion}

\NewDocumentCommand{\ibvar}{ m }{\mathsf{in}_{\mathbb{B}}(#1)}
\NewDocumentCommand{\obvar}{ m }{\mathsf{out}_{\mathbb{B}}(#1)}
\NewDocumentCommand{\ipvar}{ m }{\mathsf{in}_{\mathbb{N}}(#1)}

\NewDocumentCommand{\qrank}{  }{\mathop{\mathsf{qr}}}

\newcommand{\PSPACE}{\ensuremath{\mathsf{PSPACE}}}
\newcommand{\TOWER}{\ensuremath{\mathsf{TOWER}}}
\knowledge{notion}
 | high-level language
 | high-level for-program
 | For-program
 | (high-level) for-programs
 | high-level for-programs
 | for-programs
 | for-program

\knowledge{notion}
 | first-order Hoare triple
 | first-order Hoare triples
 | Hoare triple
 | Hoare triples

\knowledge{notion}
 | evaluation context
 | evaluation environment
\knowledge{notion}
 | list expression
 | list expressions
\knowledge{notion}
 | boolean expressions
 | boolean expression
\knowledge{notion}
 | generator expression
 | generator expressions
\knowledge{notion}
 | constant expressions
\knowledge{notion}
 | control statements
 | high-level control statements

\knowledge{notion}
 | Simple for-program
 | simple for-program
 | simple for-programs
 | first-order simple for-program
 | first-order simple for-programs
 | First-order simple for-programs

\knowledge{notion}
 | model checking problem
 | model checking algorithm
 | model checking

\knowledge{notion}
 | nested words
 | nested-word
 | (nested) words

\knowledge{notion}
 | First-order formula
 | first-order formula
 | first-order formulas
 | first-order formula on words
 | first-order logic 
 | first-order logic on words
 | first-order logic on finite words

\knowledge{notion}
 | emptiness
 | emptiness problem
 | unsatisfiable

\knowledge{notion}
 | character constants
\knowledge{notion}
 | output function
\knowledge{notion}
 | arity function
 | arity@formula
\knowledge{notion}
 | transduction tags
 | tags@transduction
\knowledge{notion}
 | domain formula
 | domain@formula
\knowledge{notion}
 | ordering formula
 | ordering formulas
 | order formula
 | order@formula

\knowledge{notion}
 | First-order string-to-string interpretations
 | first-order string-to-string interpretations
 | first-order string-to-string interpretation
 | first-order word-to-word interpretation
 | first-order interpretation
 | first-order interpretations
 | $\FO$-interpretation
 | $\FO$-interpretations
 | $\FO$ interpretation
 | $\FO$ interpretations

\knowledge{notion}
 | quantifier ranks
 | quantifier rank

\knowledge{notion}
 | loop depth
\knowledge{notion}
 | boolean depth

\knowledge{notion}
 | supported by
 | supported

\knowledge{notion}
 | touch

\knowledge{notion}
 | pullback operator
 | pullback operation

\knowledge{notion}
 | program formulas
 | program formula
 | Program formula

\knowledge{notion}
 | output boolean variable
 | output boolean variables
\knowledge{notion}
 | input boolean variable
 | input boolean variables
\knowledge{notion}
 | input position variable

\knowledge{notion}
 | regularity preserving
 | regularity preserving functions
 | regularity preserving property

\knowledge{notion}
 | linear regular functions
 | linear regular function

\knowledge{notion}
 | first-order polyregular functions
 | first-order polyregular function
 | star-free polyregular functions
\knowledge{notion}
 | polyregular functions
 | two-way deterministic transducers with nested pebbles

\knowledge{notion}
 | MONA
\knowledge{notion}
 | Z3
\knowledge{notion}
 | CVC5

\knowledge{notion}
 | Elimination of Literal Equalities
 | elimination of literals
 | elimination of Literal Equality
\knowledge{notion}
 | Elimination of Literal Productions
\knowledge{notion}
 | Elimination of Function Calls
\knowledge{notion}
 | Elimination of Boolean Generators
\knowledge{notion}
 | Elimination of Let Output Statements
\knowledge{notion}
 | Elimination of Return Statements
\knowledge{notion}
 | Defining booleans at the beginning of for loops
\knowledge{notion}
 | Expansion of For Loops
 | for loop expansion
 | expansion of for loops

\title{Polyregular Model Checking}

\author{
Aliaume Lopez\thanks{University of Warsaw, Poland. Aliaume Lopez was supported by the Polish National Science Centre (NCN) grant ``Polynomial finite state computation'' (2022/46/A/ST6/00072).}
 \and
Rafał Stefański\thanks{University of Warsaw, Poland. Rafał Stefański was supported by the European Research Council (ERC) under the European Union's Horizon 2020 innovation program (grant PROCONTRA-885666).}
}

\date{2025-04-30 11:27:16 +0200 -- 4f9e99fced29e436c1a82d50b9a762d2c3f1c612\footnote{branch main at git@github.com:AliaumeL/polyregular-model-checking.git}}

\newcommand{\repositoryUrl}{\url{https://github.com/AliaumeL/polyregular-model-checking}}

\begin{document}
\maketitle
\begin{abstract}
    We introduce a high-level language with Python-like syntax for string-to-string, polyregular, first-order definable transductions. This language features function calls, boolean variables, and nested for-loops. We devise and implement a complete decision procedure for the verification of such programs against a first-order specification. The decision procedure reduces the verification problem to the decidable first-order theory of finite words (extensively studied in automata theory), which we discharge using either complete tools specific to this theory (MONA), or to general-purpose SMT solvers (Z3, CVC5).
\end{abstract}

\section{Introduction}
\label{sec:intro}

String manipulating programs of low complexity are ubiquitous in modern
software. They are often used to transform data and do not perform complex
computations themselves. In this paper, we are interested in verifying
\kl{Hoare triples} for such string manipulating programs, i.e.
specifications of the form $\hoaretriple{P}{\texttt{code}}{Q}$, where $P$ and $Q$ are pre- and
post-conditions, meaning that whenever the input satisfies property $P$, the
output of the program satisfies property $Q$.

\paragraph{Regularity preserving programs.} \AP One particularly interesting
class of specifications in the case of string-to-string functions are
\emph{regular languages}, which can be efficiently verified using
automata-based techniques. We say that a function $f$ is \intro{regularity
preserving} if it preserves regular languages under pre-image, i.e. if
$f^{-1}(L)$ is regular for all regular languages $L$. For \kl{regularity
preserving functions}, the verification of a Hoare triple
$\hoaretriple{L_P}{f}{L_Q}$ can be reduced to the nonemptiness problem of the
language $L_P \cap f^{-1}(L_Q)$, where $L_P$ and $L_Q$ are regular languages.
This is a well-studied problem in the literature, and is at the core of several
more involved techniques \cite{ALCE11,CHLRW19,JLMR23}. The key challenge of
this approach is that there exist uncomputable \kl{regularity preserving
functions}, so such approaches will only work on classes of functions for which
pre-images of regular languages are (relatively) efficiently computable.
Usually, these classes come from generalisation of automata models to
functions, also known as \emph{string-to-string transducers}.

\paragraph{String-to-string transducer models.} There is a wide variety of
models for string-to-string transducers \cite{MUSC19}, and one of the most
prominent ones is called \intro{linear regular functions}, that are
equivalently defined using two-way finite transducers (2DFTs)
\cite{RASCO59}, streaming-string-transducers (SSTs) \cite{ALUR11}, or linear
regular list functions \cite{BDK18}. Notably, Alur and Černý have proven
that SSTs have a \PSPACE-complete model checking problem when the functions are
given as SSTs, and the specifications are given as automata
\cite[Theorem 13]{ALCE11}. This was used for instance by Chen, Taolue, Hague,
Lin, Rümer, and Wu to study \emph{path feasibility} in string-manipulating
programs \cite{CHLRW19}.

A similar approach was used by Jeż, Lin, Markgraf, and Rümmer, who leveraged
the \emph{rational functions} (a strict subclass of \kl{linear regular
functions}) to study programs manipulating strings with infinite alphabets
\cite{KAFR94}. Remark that in the setting of infinite alphabets, the landscape
of automata and transducers is much more complex: In partcicular, the class of
languages recognised by two-way automata is stronger that the class of
languages recognised by one-way automata, and has undecidable emptiness
\cite[Figure 1.1]{BOJA19}.

\AP One limitation of the \kl{linear regular functions} is that they only 
allow for linear growth of the output, excluding many useful string-manipulating 
programs. The class of \reintro{polyregular functions} is an interesting 
generalisation of \kl{linear regular functions} that allows for polynomial behaviour, 
and is much closer to real life string manipulating programs. The model 
is relatively old, first introduced in \cite{ENMA02},
and has recently gained a lot of traction now that
several other characterizations have been obtained
\cite{bojanczyk2018polyregular,bojanczyk2019string}.\footnote{Note that for this extended model, being
  \kl{regularity preserving} is tightly connected to being closed under
function composition \cite[Proposition III.3]{FIRELH25}, and this closure under
composition was one of the surprising conclusions of
\cite{bojanczyk2018polyregular}.} However, the proof of the \kl{regularity
preserving} property for polyregular functions is of theoretical nature (no
implementation or complexity bounds are given), and writing programs using any
of the existing equivalent definitions of \kl{polyregular functions} is
cumbersome and error-prone. Because \kl{polyregular functions} can succinctly
encode formulas in first-order logic on words, and
since the satisfiability problem for such formulas is known to be
\TOWER-complete \cite[Theorem 13.5]{REINH02}, one can expect that verifying
\kl{polyregular functions} to be quite complex.

\paragraph{MSO vs FO.} \AP Instead of using the full power of regular languages
(defined equivalently using finite automata, monadic second order logic ($\mathsf{MSO}$), finite monoid
recognition, or regular expressions \cite{buchi1960weak,KLEE56,TRAK66,SCHU61}),
we will use specifications written in \kl{first-order logic} ($\FO$)
on finite words. A cornerstone result of the theory is establishing the
equivalence between languages described in this logic, \emph{star-free
languages}, and \emph{counter free automata} \cite{PEPI86,SCHU65,MNPA71}. The
advantage of using this weaker specification model is twofold: first, it allows
us to focus on a simpler class of \kl{star-free polyregular
functions}\footnote{ The notion of being \emph{star-free} has been extended to
various classes of transducers, see
\cite{CADA15,BDK18,MUSC19,bojanczyk2018polyregular}}, which are easier to work
with in practice. Second, it allows us to reduce the satisfiability of a Hoare
triple to the satisfiability of a \emph{first-order} formula on finite words,
for which one can use general purpose SMT solvers, in addition to automata 
based tools (\kl{MONA}) which also work for the $\MSO$ logic on words.
Even though the SMT solver are \emph{incomplete}, 
they can, in some cases, lead to faster decision procedures. Indeed, the
satisfiability problem for \kl{first-order logic on finite words}, while
decidable, is \TOWER-complete \cite[Theorem 13.5]{REINH02}. 

\paragraph{Contributions.} \AP In this paper, we introduce a high-level
programming language for implementing \kl{star-free polyregular functions}
in a Python-like syntax, including features such as boolean variables, index variables,
immutable list variables, function calls, and nested for-loops.
The language was carefully designed not become too expressive -- 
this is ensured by a number of syntactic restrictions and a novel type system
for index variables. We show that this language can be
compiled into one of the equivalent definitions of \kl{polyregular functions}
(namely, \kl{simple for-programs}), which does not allow for function calls nor
list variables. We also provide an implementation of the previously known
abstract result stating that \kl{polyregular functions} are \kl{regularity
preserving} (in the case of star-free functions and languages), being careful
about the complexity of the transformations. Finally, we reduce the
verification of Hoare triples to the satisfiability of first-order formulas on
words. Since we are using \kl{first-order logic} as a target language, we are
not restricted to using automata based tools like \intro{MONA} \cite{MONA01},
but can also employ general purpose SMT solvers like \intro{Z3} \cite{z3} and
\intro{CVC5} \cite{cvc5}, generating proof obligations in the \texttt{SMT-LIB}
format \cite{BARRETT17}.

All the steps described above have been implemented in a \texttt{Haskell}
program, and tested on a number of examples with encouraging
results.\footnote{An anonymized version of our code is available at
\repositoryUrl.} While this is not a tool paper, we believe that the
proof-of-concept implementation is a good starting point to demonstrate the viability of
our approach, and we believe that there is a potential for further
investigations in this direction.

\paragraph{Outline.} The structure of the paper is as follows. We introduce our
\kl{high-level language} in \cref{sec:high-level}. In \cref{sec:polyregular},
we recall the theory of \kl{polyregular functions} by introducing them in terms
of \kl{simple for-programs} and \kl{$\FO$-interpretations}. We will also
provide an efficient reduction of the verification of Hoare triples to the
satisfiability of a \kl{first-order formula on words} in \cref{sec:pullback}.
In order to verify \kl{for-programs}, we compile them into \kl{simple
for-programs} in \cref{sec:htl}, and then compile \kl{simple for-programs} into
\kl{$\FO$-interpretations} in \cref{sec:low-level}. 
Then, in \cref{sec:benchmarks}, we present
tests of our implementation on various examples, discussing
the complexity of the transformations and the main bottlenecks of our approach.
Finally,  in \cref{sec:conclusion}, we discuss potential
optimizations and future work.

\section{High Level For Programs}
\label{sec:high-level}

\AP In this section, we introduce our \kl{high-level language} for
describing list-manipulating functions which can be seen as a subset of
\texttt{Python}, which we call \reintro{(high-level) for-programs}. Our goal is
to reason algorithmically about the programs written in this language, so it
needs to be highly restricted. To illustrate those restrictions, let us present
in \cref{fig:python-example-nested}
a comprehensive example written in a subset of \texttt{Python}.\footnote{The
corresponding program in the syntax accepted by our solver is given in
\cref{fig:high-level-example-nested}.}

\begin{figure}[h]
    \centering
\begin{Shaded}
\begin{Highlighting}[numbers=left]
\KeywordTok{def}\NormalTok{ getBetween(l, i, j):}
    \CommentTok{""" Get elements between i and j """}
    \ControlFlowTok{for}\NormalTok{ (k, c) }\KeywordTok{in} \BuiltInTok{enumerate}\NormalTok{(l):}
        \ControlFlowTok{if}\NormalTok{ i }\OperatorTok{\textless{}=}\NormalTok{ k }\KeywordTok{and}\NormalTok{ k }\OperatorTok{\textless{}=}\NormalTok{ j:} \circled{1}{code:comparisons}
            \ControlFlowTok{yield}\NormalTok{ c} \circled{2}{code:yield}

\KeywordTok{def}\NormalTok{ containsAB(w):}
    \CommentTok{""" Contains "ab" as a subsequence """}
\NormalTok{    seen\_a }\OperatorTok{=} \VariableTok{False} \circled{3}{code:mutbool}
    \ControlFlowTok{for}\NormalTok{ (x, c) }\KeywordTok{in} \BuiltInTok{enumerate}\NormalTok{(w):}
        \ControlFlowTok{if}\NormalTok{ c }\OperatorTok{==} \StringTok{"a"}\NormalTok{:} \circled{4}{code:string:comp}
    \NormalTok{            seen\_a }\OperatorTok{=} \VariableTok{True} \circled{5}{code:settrue} 
        \ControlFlowTok{elif}\NormalTok{ seen\_a }\KeywordTok{and}\NormalTok{ c }\OperatorTok{==} \StringTok{"b"}\NormalTok{:}
            \ControlFlowTok{return} \VariableTok{True}
    \ControlFlowTok{return} \VariableTok{False}

\KeywordTok{def}\NormalTok{ subwordsWithAB(word):}
    \CommentTok{""" Get subwords that contain "ab" """}
    \ControlFlowTok{for}\NormalTok{ (i,c) }\KeywordTok{in} \BuiltInTok{enumerate}\NormalTok{(word):} \circled{6}{code:enumerate}
        \ControlFlowTok{for}\NormalTok{ (j,d) }\KeywordTok{in}\BuiltInTok{ reversed(}\BuiltInTok{enumerate}\NormalTok{(word)):} \circled{7}{code:enumerate:rev}
\NormalTok{            s }\OperatorTok{=}\NormalTok{ getBetween(word, i, j)} \circled{8}{code:immutable:variable}
            \ControlFlowTok{if}\NormalTok{ containsAB(s):}
                \ControlFlowTok{yield}\NormalTok{ s}
\end{Highlighting}
\end{Shaded}

     \caption{A small Python program that
        outputs all subwords of a given word containing \texttt{ab}
        as a scattered subword}.
    \label{fig:python-example-nested}
\end{figure}

For the sake of readability, we implicitly coerce generators (created using the
\texttt{yield} keyword) to lists. Our programs will only deal with three kinds
of values: booleans ($\intro*\Bools$), non-negative integers ($\Nat$), and
\intro{(nested) words} ($\intro*\NestedWords$), i.e. characters
($\reintro*\NestedWords[0]$), words ($\NestedWords[1]$), lists of words
($\NestedWords[2]$), etc. 
These lists can be created by \emph{yielding} values in a loop, such
as in \circleref{2}{code:yield} of \cref{fig:python-example-nested}.
In order to ensure decidable verification of Hoare triples,\footnote{
    Using \kl{first-order logic on words} as a specification language.
} we
also will enforce the following conditions, which are satisfied in our example:
\begin{enumerate}[label=(\Roman*), ref=R. \Roman*]
    \item \textbf{Loop Constructions.}
        \label{item:loop-constructions}
        We only allow \texttt{for} loops iterating forward
        or backward over a list, as in 
        \circleref{6}{code:enumerate} and \circleref{7}{code:enumerate:rev}.
        In particular, \texttt{while} loops and recursive functions 
        are forbidden, which guarantees termination of our programs.

    \item \textbf{Mutable Variables.} 
        \label{item:mut-variables}
        The only mutable variables are booleans. The
        values of integer variables are introduced by the \texttt{for} loop
        as in \circleref{6}{code:enumerate},
        and their values are fixed during each iteration. Mutable integer
        variables could serve as unrestricted counters, resulting in
        undecidable verification. Similarly, we prohibit mutable list
        variables, as their lengths could be used as counters.
        However, we still allow the use of immutable
        list variables, as in \circleref{8}{code:immutable:variable}.

    \item \textbf{Equality Checks.}
        \label{item:equality-checks}
        We disallow equality
        checks between two \kl{nested words}, 
        unless one of them is a constant expression.
        This is what happens in point \circleref{4}{code:string:comp}
        of our \cref{fig:python-example-nested}.
        Without this restriction, verification would also be undecidable
        (\cref{lem:umc-equality-nested-words}).
        More generally, classical string \emph{algorithms}
        (edit distance, string matching, longest common subsequence, etc.) should not 
        be expressible in our language, since one can easily derive an
        equality check from them.
        
    \item \textbf{Integer Comparisons.} 
        \label{item:integer-comparisons}
        The only allowed operations on integers
        are usual comparisons operators (equality, inequalities).
        However, we only
        allow comparisons between integers that are indices of the
        same list.
        Every integer is associated to a list expression.
        For instance, in points \circleref{6}{code:enumerate} and
        \circleref{7}{code:enumerate:rev} of our example, the variables
        $i$ and $j$ are associated to the same list variable \texttt{word}.
        Similarly, for the comparison 
        of point \circleref{1}{code:comparisons} to be valid,
        the variables $k$, $i$, and $j$ should all be associated to the same 
        list variable $l$.

        To ensure this compatibility, we designed the following type system,
        containing Booleans, \kl{nested words} of a given depth
        (characters are of depth $0$), and integers associated to a \kl{list
        expression} (the set of which is denoted by $\oexpr$, and will
        be defined in \cref{fig:out-expr}):
        \begin{align*}
            \tau ::=~ \intro*\TBool
            ~\mid~ \intro*\TPos[o] 
            ~\mid~ \intro*\TOut[n] 
            \quad 
            n \in \Nat, \,
            o \in \oexpr
            \quad .
        \end{align*}
        These types can be inferred from the context,
        except in the case of function arguments, in which case
        we explicitly specify to which list argument integer variables
        are associated, as shown in \cref{fig:high-level-example-nested}.

        Without this restriction, the equality predicate between two lists can
        be redefined (as shown in \cref{fig:eq-def-different-indices}).

    \item \textbf{Variable Shadowing.} 
        \label{item:variable-shadowing}
          We disallow shadowing of variable names, as it could
          be used to forge the origin of integers, leading to unrestricted comparisons
          (as shown in \cref{fig:eq-def-shadowing}).

    \item \textbf{Boolean Arguments.}
        \label{item:boolean-arguments}
        We disallow functions to take boolean arguments,
        as it
        would allow to forge the origin of integers,
        by considering the function \texttt{switch(b, l1, l2)} which
        returns either \texttt{l1} or \texttt{l2} 
        depending on the value of \texttt{b}
        (as in \cref{fig:eq-def-boolean}).

    \item \textbf{Boolean Updates.} 
        \label{item:boolean-updates}
        Boolean variables are initialized to \texttt{false}
        as in \circleref{3}{code:mutbool}, and
        once they are set to \texttt{true} as in 
        \circleref{5}{code:settrue},
        they cannot be reset to \texttt{false}. 
        We depart here from the semantics of Python by
        considering lexical scoping of variables; in
        particular a variable declared in a loop is not
        accessible outside this loop.

        This restriction allows us to reduce the verification problem to
        the satisfiability of a \kl{first-order formula} on finite words. This
        problem is not only decidable but also solvable by well-engineered
        existing tools, such as automata-based solvers (e.g., \kl{MONA}) and
        classical SMT solvers (e.g., \kl{Z3}, and \kl{CVC5}).
        Without this restriction, the problem would require the use the monadic
        second order logic on words which is still decidable but not supported
        by the SMT solvers. 

\end{enumerate}

\paragraph{Formal Syntax and Typing.} We extend the typing system to functions
by grouping input positions with the list they are associated to. For instance,
the function \texttt{getBetweenIndicesBeforeStop(l, i, j)} has type
$(\TOut[2],2) \to \TOut[2]$, that is, we are given an input list $l$ together
with two pointers to indices of this list. Similarly, the function
\texttt{containsAB(w)} has type $(\TOut[1], 0) \to \TBool$, while the function
\texttt{subwordsWithAB(word)} has type $(\TOut[1], 0) \to \TOut[2]$. We
implemented a linear-time algorithm for the type checking and inference
problems.

\AP The formal syntax of our language is given in
\cref{fig:bool-expr,fig:const-expr,fig:out-expr,fig:high-level-stmt,fig:high-level-program}.
They define the syntax of \intro{boolean expressions} ($\bexpr$),
\intro{constant expressions} ($\cexpr$), \intro{list expressions} ($\oexpr$),
and \intro{control statements} ($\hlstmt$). For readability, we distinguish
boolean variables $\intro*\BVars$ ($b, p, q, \dots$), position variables
$\intro*\PVars$ ($i,j, \dots$), list variables $\intro*\OVars$ ($x,y,u,v,w,
\dots$), and function variables $\intro*\FunVars$ ($f,g,h, \dots$). A
\intro{for-program} is a list of function definitions together with a
\emph{main} function of type $\TOut[1] \to \TOut[1]$.

\paragraph{Semantics.} \AP Given an \intro{evaluation context} that assigns
values (functions, positions, booleans, nested words) to variables, the
semantics of \kl{boolean expressions} into booleans $\Bools$, \kl{constant
expressions} into \kl{nested words}, and \kl{list expressions} into \kl{nested
words} pose no difficulty.
For the \kl{control statements}, there is a crucial design choice regarding
the semantics of backward iteration.
While the semantics of forward iteration is unambiguous, the backward iteration 
$\hlforRev{(i,x)}{l}{s}$
could be understood in two different ways,
provided that $l$ evaluates to a list $[x_0, \dots, x_k]$:
\begin{itemize}
    \item Executing the statement $s$ for pairs
        $(k, x_k), \dots, (0, x_0)$. This corresponds to
        Python's \texttt{for (i,x) in reversed(enumerate(l))}
        ;
    \item Executing the statement $s$ for pairs
        $(0, x_k), \dots, (k, x_0)$.
        This corresponds to
        Python's \texttt{for (i,x) in enumerate(reversed(l))}.
\end{itemize}
As shown in our example program,
we use the first interpretation (see \circleref{7}{code:enumerate:rev}). In fact,
the second interpretation would allow us to define the equality predicate
between two lists, leading to undecidable verification.
\section{Polyregular Functions}
\label{sec:polyregular}

To obtain a decision procedure for the verification of \kl{Hoare triples} for
\kl{for-programs}, we will 
prove that they can be compiled to \intro{first-order polyregular functions} -- 
a class of transductions introduced in \cite{bojanczyk2018polyregular} whose model checking 
problem is decidable \cite[Theorem~1.7]{bojanczyk2018polyregular}. 
We provide two equivalent definitions of the \kl{first-order
polyregular functions}: one using \kl{first-order simple for-programs} \cite[p.
19]{bojanczyk2018polyregular} and one using the logical model of
\kl{first-order string-to-string interpretations} \cite[Definition
4]{bojanczyk2019string}, the equivalence of which 
was proven in
\cite{bojanczyk2018polyregular}.

To make the models more suitable for large alphabets (such as the Unicode
characters), we present them in a symbolic setting (which uses a simplified
version of the ideas presented in \cite{d2017power} or in
\cite[Section~3.1]{bojanczyk2023growth}). This will dramatically reduce the
size of the \kl{first-order string-to-string interpretations}, and in turn, of
the \kl{first-order formula} that we will feed to the solvers. We will prove in
\cref{sec:low-level} that every \kl{first-order simple for-programs} can be
transformed into a \kl{first-order string-to-string interpretation} in the
symbolic setting. We believe that the other inclusion should also hold,
but do not prove it, as it is out of this paper's scope.

\subsection{Symbolic transductions}

\AP
Consider the program in \cref{fig:swapAsToBs}, which swaps all \texttt{a}s to
\texttt{b}s in a string. Even though it operates on the entire Unicode
alphabet, it only distinguishes between three types of characters: \texttt{a},
\texttt{b} and the rest.
To formalize this observation, we model the Unicode alphabet as an infinite set $\mathcal{D}$, and
we define a function $T : \mathcal{D}^* \to \mathcal{D}^*$ to be
\intro{supported by} a set $A \subseteq \mathcal{D}$, if for every function $f:
\mathcal{D} \to \mathcal{D}$ that does not \intro{touch} elements of $A$ (i.e.
$\forall_{a\in A}, f^{-1}(a) = \{a\}$), it holds that:
\begin{equation*}
    \forall_{w} \quad T(f^*(w)) = f^*(T(w)) \quad ,
\end{equation*}
Where $f^*$ is the extension of $f$ to $\mathcal{D}^*$, defined by applying $f$
to every letter.
\begin{wrapfigure}{r}{0.5\textwidth}
\begin{Shaded}
\begin{Highlighting}[numbers=left]
\KeywordTok{def}\NormalTok{ asToBs(w): }
    \ControlFlowTok{for}\NormalTok{ (i, c) }\KeywordTok{in} \BuiltInTok{enumerate}\NormalTok{(w):}
        \ControlFlowTok{if}\NormalTok{ c }\OperatorTok{==} \StringTok{'a'}\NormalTok{: }
            \ControlFlowTok{yield} \StringTok{'b'}
        \ControlFlowTok{else}\NormalTok{:}
            \ControlFlowTok{yield}\NormalTok{ c}
\end{Highlighting}
\end{Shaded}
     \caption{The \texttt{swapAsToBs} program.}
    \label{fig:swapAsToBs}
\end{wrapfigure}

Functions defined by
\kl{for-programs} (of type $\TOut[1] \to \TOut[1]$)
are supported by the finite set $A$ of letter constants that they use.
This is also going to be the case for the \kl{simple for-programs}
that we introduce in \cref{subsec:simple-for-programs}.
In
\cref{subsec:fo-string-to-string}, we will define a version of the
\kl{first-order string-to-string interpretations} in a way that only depends on
the size of their support $A$, and not on the number of the Unicode characters.

\subsection{First-Order Simple For-Programs}
\label{subsec:simple-for-programs}

\AP \intro{First-order simple for-programs}
 ---  originally introduced in \cite[p.
19]{bojanczyk2018polyregular} ---
can be seen as
simplified\footnote{Actually, the \kl{for-programs} were designed as an
extended version \kl{first-order simple for-programs}.} 
version of the \kl{for-programs}. The main difference is that the \kl{simple
for-programs} only define transductions of type $\TOut[1] \to \TOut[1]$. Here
is an example in a Python syntax:

\begin{Shaded}
\begin{Highlighting}[numbers=left]
\CommentTok{\# The program reverses all space{-}separated words }
\CommentTok{\# in the input string. e.g }
\CommentTok{\#        "hello world" {-}\textgreater{} "olleh dlrow"}
\NormalTok{seen\_space\_top }\OperatorTok{=} \VariableTok{False} \circled{1}{sfp:boolDeclTop}
\CommentTok{\# first we handle all words except of the final one }
\ControlFlowTok{for}\NormalTok{ i }\KeywordTok{in} \BuiltInTok{input}\NormalTok{:} \circled{2}{sfp:for1}
\NormalTok{    seen\_space }\OperatorTok{=} \VariableTok{False} \circled{3}{sfp:boolDeclFor}
    \ControlFlowTok{if}\NormalTok{ label(i) }\OperatorTok{==} \StringTok{' '}\NormalTok{:} \circled{4}{sfp:labelTest}
        \ControlFlowTok{for}\NormalTok{ j }\KeywordTok{in} \BuiltInTok{reversed}\NormalTok{(}\BuiltInTok{input}\NormalTok{):} \circled{5}{sfp:for2}
            \ControlFlowTok{if}\NormalTok{ j }\OperatorTok{\textless{}}\NormalTok{ i: }
                \ControlFlowTok{if}\NormalTok{ label(j) }\OperatorTok{==} \StringTok{' '}\NormalTok{: }
\NormalTok{                    seen\_space }\OperatorTok{=} \VariableTok{True} 
                \ControlFlowTok{if} \KeywordTok{not}\NormalTok{ seen\_space: }
                    \BuiltInTok{print}\NormalTok{(label(j))} \circled{6}{sfp:labelPrint}
        \BuiltInTok{print}\NormalTok{(}\StringTok{' '}\NormalTok{)} \circled{7}{sfp:constPrint}

\CommentTok{\# then we handle the final word}
\ControlFlowTok{for}\NormalTok{ j }\KeywordTok{in} \BuiltInTok{reversed}\NormalTok{(}\BuiltInTok{input}\NormalTok{):} 
    \ControlFlowTok{if}\NormalTok{ label(j) }\OperatorTok{==} \StringTok{' '}\NormalTok{: }
\NormalTok{        seen\_space\_top }\OperatorTok{=} \VariableTok{True} 
    \ControlFlowTok{if} \KeywordTok{not}\NormalTok{ seen\_space\_top: }
        \BuiltInTok{print}\NormalTok{(label(j))}
\end{Highlighting}
\end{Shaded}

We disallow constructing intermediate word-values, there are no variables of
type $\TOut[n]$ for any $n$, and it is not possible to define functions (other
than the main function). As a consequence, the
for-loops can only iterate over the positions of the input word as in
\circleref{1}{sfp:for1} and \circleref{2}{sfp:for2}.
The character at a given position can be accessed using the keyword $\mathsf{label}$, 
whether when testing it (\circleref{5}{sfp:labelTest}) or when printing it in
(\circleref{6}{sfp:labelPrint}).
As we are considering a restriction of \kl{for-programs}, we only allow comparing labels
to constant characters (\ref{item:equality-checks}).
Finally, we only allow introducing boolean
variables at the top of the program (\circleref{1}{sfp:boolDeclTop}) or at the
beginning of a for loop (\circleref{3}{sfp:boolDeclFor}). 

\subsection{First-Order String-To-String Transductions}
\label{subsec:fo-string-to-string}
\AP \kl{First-order string-to-string interpretations} forms an other model 
that defines functions $\mathcal{D}^* \to \mathcal{D}^*$. 
It is based on the \intro{first-order logic on words} ($\intro*\FO$),
the syntax of which we 
recall in \cref{fig:fo-syntax}.
To evaluate such a formula $\varphi$ on a word $w \in \mathcal{D}^*$ we perform the quantifications
over the positions in $w$. The predicates $x = y$ and $x < y$ have the natural 
meaning, and $x =_L \mathtt{a}$ is checks if the $x$-th letter of $w$ is equal to $\mathtt{a}$.
Let us recall that the \intro{quantifier rank} of a formula
is the maximal number of nested quantifications in it.
\begin{figure}
    \centering
    \begin{align*}
        \varphi,\ \psi :=&~ \forall_x\; \varphi ~|~ \exists_x\; \varphi ~|~ \varphi \wedge \psi ~|~ \varphi \vee \psi ~|~ \neg \varphi \\
                |&~ x = y ~|~ x < y ~|~ x =_L \mathtt{a} \textrm{, where } \mathtt{a} \in \mathcal{D}
    \end{align*}
    \caption{First-order logic on words.}
    \label{fig:fo-syntax}
\end{figure}

\AP An important property of $\FO$, is that it has decidable \intro{emptiness},
i.e. given a formula $\varphi$, one can decide if there is a word $w$ such that
$\varphi$ holds for $w$. For finite alphabets, this property is well-know
\cite{buchi1960weak}, and for the infinite alphabet $\mathcal{D}$ it is the
consequence of the finite-alphabet case (\cref{lem:fo-emptiness}).

Having discussed the \kl{first-order logic on words}, we are now ready to
define the \kl{first-order string-to-string interpretations}.
\begin{definition}
A \intro{first-order string-to-string interpretation} consists of:
\begin{enumerate}
\item A finite set of \intro{character constants} $A \subset_{\textrm{fin}}\mathcal{D}$.
\item A finite set $T$ of \intro(transduction){tags}.
\item An \intro{arity function} $\mathsf{ar} : T \to \Nat$.
\item An \intro{output function} $\mathsf{out} : T \to A + \{1, \ldots, \mathsf{ar}(t)\}$. 
\item A \intro{domain formula} $\varphi_{\mathsf{dom}}^t(x_1,\ldots,x_{\mathsf{ar}(t)})$ 
    for every tag $t \in T$.
\item An \intro{order formula} $\varphi_{\leq}^{t,t'}(x_1,\ldots,x_{\mathsf{ar}(t)},y_1,\ldots,y_{\mathsf{ar}(t')})$ for every $t,t' \in T$. 
\end{enumerate}
The \kl(formula){order} and \kl(formula){domain} formulas should only use constants from $A$.
\end{definition}
\noindent
The interpretation's output for a word $w \in \mathcal{D}^*$ is obtained as follows:
\begin{enumerate}
    \item  Take the set $P = \{1, \ldots, |w|\}$ of the positions in $w$, and construct the set 
           of \emph{elements} as the set $T(P) = (t : T) \times P^{\mathsf{ar}(t)}$
           of all tags from $T$ equipped with position tuples of the appropriate arity.
    \item Filter out the elements that do not satisfy the domain formula.
    \item Sort the remaining elements according to the order formula. Typically, we 
          want the order formula to define a total order on the remaining elements of $T(P)$ -- 
          if this is not the case, the interpretation returns an empty word.
    \item Assign a letter to each element according to the output function: For an 
          element $t(p_1, \ldots, p_k)$, we look at of $\text{out}(t)$: If it returns $a \in A$
          the output letter is $a$. If it returns $i \in \{1, \ldots, k\}$, we copy the output letter from the
          $p_i$-th position of the input.
          
\end{enumerate}

\begin{wrapfigure}{r}{0.5\textwidth}
\[
\begin{tabular}{ccc}
    $\text{out}(\mathtt{printB}) = \mathtt{b}$ & \ \ \ & $\text{out}(\mathtt{copy}) = 1$ \\
\end{tabular}
\]
\[
\begin{tabular}{ccc}
    $\varphi_{\text{dom}}^{\mathtt{printB}}(x) : x =_L \mathtt{b}$ & \ \ \  &$\varphi_{\text{dom}}^{\mathtt{copy}}(x) : x \neq_L \mathtt{b}$ \\
\end{tabular}
\]
\[ 
\begin{tabular}{c|cc}
    $\varphi_{\leq}$ & $\mathtt{printB}(x_1)$ & $\mathtt{copy}(x_1)$ \\
    \hline
    $\mathtt{printB}(x_2)$ & $x_1 \leq x_2$ & $x_1 < x_2$ \\
    $\mathtt{copy}(x_2)$ & $x_1 \leq x_2$ & $x_1 \leq x_2$ \\
\end{tabular}
\]
\caption{The \texttt{swapAsToBs} interpretation.}
\label{fig:swapAsToBsInterpretation}
\end{wrapfigure}
For example, let us present a \kl{first-order word-to-word interpretation} for
the function \texttt{swapAsToBs}
in \cref{fig:swapAsToBsInterpretation}. It has two tags \texttt{printB} and
\texttt{copy}, both of arity~$1$. The element $\mathtt{printB}(x)$ outputs the
letter \texttt{b} and $\mathtt{copy}(x)$ outputs the letter of $x$-th position
of the input word.
The element $\mathtt{printB}(x)$ is present in the output if $x$ is labelled with the letter \texttt{b}
in the input, otherwise the element $\mathtt{copy}(x)$ is present:
The tags are sorted by their positions, with ties resolved in favour of \texttt{printB}.

\subsection{Hoare Triple Verification}
\label{sec:pullback}

\AP
We say that the Hoare triple $\{ \varphi \}\; F\; \{ \psi \}$ is valid if
for every word $w$ that satisfies $\varphi$, the output $F(w)$ satisfies
$\psi$. An important property \kl{first-order string-to-string interpretations}
is that they admit a direct reduction of the \intro{first-order Hoare triple}
verification problem to the \kl{emptiness problem} for the \kl{first-order
logic on words} \cite[Theorem~1.7]{bojanczyk2018polyregular}. However,
the resulting construction is not efficient.
We provide a direct construction of a first-order formula $\chi(\phi, F, \psi)$
that is \kl{unsatisfiable} if and only if the triple $\{ \phi \}\; F\; \{ \psi \}$ is valid.
Moreover, the size and the \kl{quantifier rank} of $\chi$ are bounded by the following low-degree polynomials:
\begin{align*}
    \qrank(\chi) \leq \max\left(\qrank(\phi),\ \qrank(\psi) \cdot \left(\textsf{ar}(F) + 1\right)  + \qrank(F)\right) \quad |\chi| = \mathcal{O} (|\phi| + |F| \cdot |\psi|)
\end{align*}

Here $|F|$ denotes the sum of the sizes of formulas in $F$,
$\qrank(F)$ denotes quantifier depth of the deepest formula in $F$, 
and $\textsf{ar}(F)$ denotes the maximal arity of the tags in $F$.

To construct the formula $\chi$, we introduce a \intro{pullback operator}
$\pi(F, \psi)$ that transforms the formula $\psi$ applied to the output $F$, to
a formula $\pi(F, \psi)$ that can be applied directly the input word,
corresponding to a form of \emph{weakest precondition} \cite[Chapter
7]{WINSKEL93}. The pull-back operation is defined in such a way that $F(w)$
satisfies $\psi$ if and only if $w$ satisfies $\pi(F, \psi)$. Once we have the
pull-back operation, we can define $\chi(\phi, F, \psi)$ as $\phi \wedge \neg
\pi(F, \psi)$. In the rest of this section, we show how to efficiently
construct $\pi(F, \psi)$.

\paragraph{Naïve Pullback Definition.} Let us start 
with a simple but inefficient construction of the \kl{pullback operation}.
Every position from $F(w)$ corresponds to a tag $t$ and a tuple of $\mathsf{ar}(t)$ positions of the input word $w$, 
so we can replace each quantification in $\psi$ with a conjunction or disjunction
over the tags,
and use respectively the \kl{order formula}
and \kl{output function} to implement the
predicates over positions of $F(w)$.
For example: 
\[ 
\begin{tabular}{ccc}
    $\forall_{x}, \psi$ & $\quad \rightsquigarrow \quad$ & $\bigwedge_{t_x \in T} \forall_{x_1, \ldots, x_{\mathsf{ar}(t)}} \left( \varphi_{\mathsf{dom}}^t(x_1, \ldots, x_{\mathsf{ar}(t)}) \Rightarrow \psi'_{t} \right)$
\end{tabular}
\]
A similar transformation can be done for the existential
quantifications. Then, one can
implement the $\leq$ predicate by consulting the \kl{order formula}:
\[
\begin{tabular}{ccc}
    $x \leq y$ & $\quad \rightsquigarrow \quad$ & $\psi_{\leq}^{t_x, t_y}(x_1, \ldots, x_{\mathsf{ar}(t_x)}, y_1, \ldots, y_{\mathsf{ar}(t_y)})$
\end{tabular}
\]
Similarly the $=_L$ predicate can be handled by consulting the \kl{output function}, and
$x = y$ predicate can be handled by comparing equality of the \kl(transduction){tags}
and the positions of $x$ and $y$. This construction, although correct, is unfortunately inefficient:  Replacing each quantification with a disjunction or conjunction
over tags, results in an exponential blow-up of the formula. 

\paragraph{Efficient Pullback Definition.} Let us introduce an additional
finite sort $T$ to the logic, which allows us to quantify over the \kl(transduction){tags} using
$\forall_{t \in T} \varphi$ and $\exists_{t \in T} \varphi$. This does not add
expressive power to the logic, as the new quantifiers can be replaced by a
finite conjunction (resp. disjunction) that goes through the tags. However,
this new sort will allow us to construct the \kl{pullback operator} in a more
efficient way, that can be understood by the solvers (we discuss it in more
details at the end of this section). With the new sort of \kl(transduction){tags}, we can
pull back the quantifiers in the following way: 
\[ 
\begin{tabular}{ccc}
   $ \pi(F, \forall_{x} \psi)$ & $\quad = \quad$ & $\forall_{t_x \in T} \; \forall_{x_1, \ldots, x_{\mathsf{ar}(F)}} \left( {\mathsf{dom}}(t_x, x_1, \ldots, x_{\mathsf{ar}(F)}) \Rightarrow \pi(F, \psi)\right)$
\end{tabular}
\]
Where $\mathsf{dom}$ is the following predicate based on 
the \kl{domain formula}:
\begin{align*}
    \mathsf{dom}(t, x_1, \ldots, x_{\mathsf{ar}(t)}) :=&~ \bigvee_{t' \in T} \left( t = t' \wedge \varphi_{\mathsf{dom}}^{t'}(x_1, \ldots, x_{\mathsf{ar}(t')}) \right)
\end{align*}
In order to implement the atomic predicates,
we use  formulas similar to $\mathsf{dom}$, but based the \kl{order formula} and \kl{output function}:
\begin{align*}
   \pi(F, x \leq y)  =&~ \bigvee_{t_1, t_2 \in T} \left( t_x = t_1 \wedge t_y = t_2 \wedge \varphi_{\leq}^{t_1, t_2}(x_1, \ldots, x_{\mathsf{ar}(t_1)}, y_1, \ldots, y_{\mathsf{ar}(t_2)}) \right) \\
   \pi(F, x =_L \mathtt{a}) =&~ \left(\bigvee_{t \in T \wedge \mathsf{out}(t) = \mathtt{a}} t = t_x\right) \vee
\left(\bigvee_{t \in T \wedge \mathsf{out}(t) \not \in A} (t = t_x \wedge x_{\mathtt{out}(t)} =_L \mathtt{a})\right)
\end{align*}
This way, we push the disjunction over \kl(transduction){tags} all the way down in the formula,
thus avoiding the exponential blow-up of the naïve approach.

\paragraph{Encoding tags.} Finally, let us briefly discuss how we handle the
tags in the formulas fed to solvers. For the SMT-solvers, we use the
\texttt{smtlib v2.6} format with logic set to \texttt{UFDTLIA}
\cite{BARRETT17}, which allows us to add finite sorts and quantify over them.
For the MONA solver, which only supports the sort of positions, we encode the
tags as the first $|T|$ positions of the input word. The pertinence of
this choice of encoding will be discussed in \cref{sec:benchmarks}.
\section{From High Level to Low Level For Programs}
\label{sec:htl}

\AP In this section, we provide a compilation from \kl{high-level for-programs}
to \kl{simple for-programs}. To smoothen the conversion, we introduce
\intro{generator expressions} to the language, as a way to inline 
function calls. We distinguish between \kl{nested-word} generators $\intro*\ogen{s}$
and boolean generators $\intro*\bgen{s}$.

\paragraph{Generator Expressions.} Let us briefly discuss the new typing rules
and semantics of these \kl{generator expressions}. The meaning of 
$\ogen{s}$ is to evaluate the statement $s$ in the current context and collect its output.
For instance, $\ogen{\hlreturn{x}}$ is equivalent to $x$,
and $\ogen{\hlseq{\hlyield{x}}{\hlyield{y}}}$ is equivalent
to $\olist{x,y}$. Similarly, $\bgen{s}$ is used to evaluate a boolean
statement and return its value. The type of a \kl{generator expression}
is equal to the type of the statement $s$ it contains. Importantly, 
when evaluating the statement $s$ in a generator,
we hide all boolean variables from the \kl{evaluation context}.
In particular, $\hlletboolean{b}{\hlreturn{\bgen{ \hlreturn{b} }}}$ is an \emph{invalid
program}, because the variable $b$ is undefined in the context of the generator
expression $\bgen{b}$. The formal typing rules of \kl{generator expressions}
can be found in \cref{fig:generators}. 

Hiding the booleans from the context, ensures that the evaluation order of 
the expressions is irrelevant, allowing us to freely substitute expressions
during the compilation process. 

\paragraph{Rewriting Steps.} We will convert \kl{for-programs} to
\kl{simple for-programs} by a series of rewriting steps listed below.
While most of the steps make can be applied to any \kl{for-program},
some of them only apply to programs of type $\TOut[1] \to \TOut[1]$.
\begin{enumerate}[label=(\Alph*), ref=Step \Alph*]
    \item \label{item:lit_eq_elim} \intro{Elimination of Literal
        Equalities}, i.e., of expressions $\bliteq{c}{o}$ where $c \in \cexpr$
        and $o \in \oexpr$. This is done by replacing those tests with a call
        to a function that checks for equality with the constant $c$ by
        traversing its input. We define these functions by induction on $c$.
        Note that this is only possible because equalities are always
        between a variable and a constant (\ref{item:equality-checks}).

    \item \label{item:lit_elim} \intro{Elimination of Literal
        Productions}, i.e., of constant expressions in the construction of
        $\oexpr$, except single characters. This is done by replacing a
        constant $c$ by a function call. For instance, $\clist{\cchar{a_1}, \cchar{a_2}}$
        is replaced by a call to a function with body
        $\hlseq{\hlyield{\cchar{a_1}}}{\hlyield{\cchar{a_2}}}$.

    \item \label{item:fun_elim} \intro{Elimination of Function Calls},
        by replacing them with \kl{generator expressions}. Given a function $f$
        with body $s$ and arguments $x_1, \dots, x_n$, we replace a call
        $f(a_1, \dots, a_n)$ by $\ogen{ s[ a_1/x_1, \dots, a_n/x_n ] }$
        (or $\bgen{ \cdots }$ for boolean functions).
        This is valid because functions do not take booleans as arguments
        (\ref{item:boolean-arguments}).

    \item \label{item:bool_elim} \intro{Elimination of Boolean
        Generators}. Note that $\bgen{s}$ can only appear in a conditional test, 
        and let us illustrate this step on an example.
        Consider the following statement:
        $\hlif{\bgen{s_1}}{s_2}{s_3}$. We replace it by $\hlletboolean{b_1}{
        (\hlseq{s_1'}{\hlif{b_1}{s_2}{s_3}}) }$, where $s_1'$ is obtained by
        replacing boolean return statements $(\hlreturn{b})$ by assignments of
        the form $(\hlif{b}{\hlsettrue{b_1}}{\hlskip{}})$.
    \item \label{item:let_output_elim} \intro{Elimination of Let
        Output Statements}, i.e., of statements of the form
        $\hlletoutput{x}{e}{s}$. This is done by textually replacing
        $\hlletoutput{x}{e}{s}$ by $s[x \mapsto e]$.

    \item \label{item:return_elim} \intro{Elimination of Return
        Statements} for \kl{list expressions}. 
        First, to make sure that the program does not produce 
        any output after the first return statement, we
        introduce a boolean variable \texttt{has\_returned}, 
        and guard every yield statement by a check on this variable.
        Then, we replace every statement $\hlreturn{e}$ by
        a for loop $\hlfor{(i,x)}{e}{\hlyield{x}}$.
        This is not possible if the return statement is of type $\TOut[0]$,
        and for this edge case, we refer the readers to our implementation.

    \item \label{item:for_loop_exp} \intro{Expansion of For Loops},
        ensuring that every for loop iterates over a single list
        variable. This is the key step of the compilation, and it will be 
        thoroughly explained later in this section.

    \item \label{item:let_bools_top} 
        \intro{Defining booleans at the beginning of for loops}.
        This is a technical step that ensures that all boolean variables 
        are defined at the beginning of the program or at the beginning of a for loop.
        Thanks to the no-shadowing rule (\ref{item:variable-shadowing}), we can 
        safely move all boolean definitions to the top of their scopes.
\end{enumerate}

\begin{theorem}
    \label{thm:rewriting-termination}
    The rewriting steps (\ref{item:lit_eq_elim}--- \ref{item:let_bools_top})
    all terminate and preserve typing. Moreover, normalized \kl{for-programs}
    of type $ \TOut[1] \to \TOut[1]$ are isomorphic to \kl{simple for-programs}.
\end{theorem}

\paragraph{Forward For Loop Expansion.} We now focus on the \kl{expansion of for
loops}, that is, \ref{item:for_loop_exp}. The case of forward
iterations is simpler and will illustrate a first difficulty. We replace each loop of
the form $\hlfor{(i,x)}{\ogen{s_1}}{s_2}$ by the statement $s_1$ where every
statement $\hlyield{e}$ is replaced by $s_2[x \mapsto e]$. This
rewriting is problematic because it leaves the variable $i$ undefined in $s_2$.
The key observation allowing us to circumvent this issue is that the variable
$i$ can only be used in \emph{comparisons}, and can only be compared with 
variables $j$ that are iterating over $\ogen{s_1}$
(thanks to \ref{item:integer-comparisons}). It is therefore sufficient 
to order the outputs of $s_1$ to effectively remove the variable $i$ from the program.

One can recover the ordering between outputs of $s_1$ by storing
the position of the $\hlyield{e}$ responsible for the output,
together with all position variables visible at that point.
Let us illustrate this in a simple example:
\begin{equation*}
    \hlseq{
    (\hlforRev{(j\tikzmark{yieldIndex},y)}{e}{
        (\hlseq{\tikzmark{yield1}\hlyield{y}}
               {\tikzmark{yield2}\hlyield{\cchar{a}}}
        )})
    }{\tikzmark{yield3}\hlyield{\cchar{b}}}
\end{equation*}
\begin{tikzpicture}[overlay, remember picture]
    \node (Y1) at ([yshift=-0.5cm, xshift=0.2cm]pic cs:yield1) {$p_1(j)$};
    \node (Y2) at ([yshift=-0.5cm, xshift=0.2cm]pic cs:yield2) {$p_2(j)$};
    \node (Y3) at ([yshift=-0.5cm, xshift=0.2cm]pic cs:yield3) {$p_3$};
    \node (YI) at ([yshift=0cm, xshift=-0.2cm]pic cs:yieldIndex) {};

    \draw[dashed, A2, thick] (Y1) edge[->, bend left] (YI);
    \draw[dashed, A2, thick] (Y2) edge[->, bend left] (YI);
\end{tikzpicture}
\vspace{1em}

In this example, there are three yield statements at
positions $p_1$, $p_2$ and $p_3$. We can compute
the \emph{happens (strictly) before} relation between outputs 
of the various yield statements:
\begin{center}
    $\mathsf{before}(p_1(j), p_2(j)) = \btrue$ \hspace{1em}
    $\mathsf{before}(p_2(j), p_3) = \btrue$ \hspace{1em}
    $\mathsf{before}(p_1(j), p_3) = \btrue$ 
    \\[1em]
    $\mathsf{before}(p_1(j), p_1(j')) = j > j'$ \hspace{2em}
    $\mathsf{before}(p_2(j), p_2(j')) = j > j'$
    \\[1em]
    $\mathsf{before}(p_1(j), p_2(j')) = j \geq j'$
\end{center}
In the case of $j = j'$, the 
output of $p_1(j)$ happens before the output of $p_2(j')$,
because $p_1$ is the first yield statement in the loop.
When $j > j'$, the output of $p_1(j)$ happens
before the output of $p_2(j')$ because the loop
is iterating in reverse order.

\paragraph{Backward For Loop Expansion.} The case of backward iterations adds a
new layer of complexity, namely to perform a non-reversible computation $s$ in
a reversed order: indeed, in the for loop $\hlforRev{(i,x)}{\ogen{s_1}}{s_2}$,
$s_1$ can contain the command $\hlsettrue{b}$ which cannot be reversed.

Let us consider as an example
$\hlforRev{(i,x)}{\ogen{s}}{\hlyield{x}}$, where the statement $s$ is defined
to print all elements of a list $u$ except the first one, namely:
\[ s \ :=\ \hlletboolean{b}{\hlfor{(j,y)}{ u }{\hlif{b}{\hlyield{y}}{\hlsettrue{b}}}}\] 
The semantics of $\hlforRev{(i,x)}{\ogen{s}}{\hlyield{x}}$ is to print all
elements of $u$ in reverse order, skipping the last loop iteration. To compute
this new statement, we will use the following \emph{trick} that can be traced
back to \cite[Lemma 8.1 and Figure 6, p. 68]{bojanczyk2018polyregular}: we will
use two versions of the statement $s$, the first one $s_\mathsf{rev}$, will be
$s$ where all boolean introductions are removed, if statements
$\hlif{e}{s_1}{s_2}$ are replaced by sequences $\hlseq{s_1}{s_2}$, every loop
direction is swapped, and every sequence of statements is reversed. Its
intended semantics is to reach all possible yield statements of $s$ in the
reversed order. In our case:
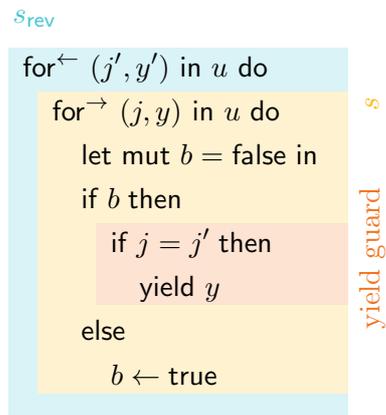
\begin{wrapfigure}{r}{0.5\textwidth}
    \centering
\begin{tikzpicture}[overlay, remember picture]
    \draw[fill=D4hint,draw=none] 
                     ([yshift=1em, xshift=-0.5em]pic cs:revloopexpstart) 
                     rectangle 
                     ([yshift=-1.3em, xshift=4.5em]pic cs:revloopexpend);

    \draw[fill=D3hint,draw=none] 
                     ([yshift=1em, xshift=-0.5em]pic cs:revloopexpinnerstart) 
                     rectangle 
                     ([yshift=-0.3em, xshift=4.5em]pic cs:revloopexpend);

    \draw[fill=D2hint,draw=none] 
                     ([yshift=1em, xshift=-0.5em]pic cs:revloopifstart)
                     rectangle 
                     ([yshift=-0.3em, xshift=4.4em]pic cs:revloopifend);

     \node[anchor=east,D4] at ([xshift=1.5em,yshift=2em]pic cs:revloopexpstart) {$s_{\mathsf{rev}}$};
     \node[anchor=west,D3,rotate=90] at ([xshift=11em]pic cs:revloopexpinnerstart) {$s$};
     \node[anchor=west,D2,rotate=90] at ([xshift=9em,yshift=-3em]pic cs:revloopifstart) {yield guard};
\end{tikzpicture}
\begin{align*}
    &\tikzmark{revloopexpstart}\mathsf{for}^{\leftarrow}~(j',y')~\mathsf{in}~ u ~\mathsf{do} \\
    &\quad \tikzmark{revloopexpinnerstart}\mathsf{for}^{\rightarrow}~(j, y)~\mathsf{in}~ u ~\mathsf{do} \\
    &\quad \quad \mathsf{let}~\mathsf{mut}~b = \mathsf{false}~\mathsf{in} \\
    &\quad \quad \mathsf{if} \; b \; \mathsf{then} \\
    &\quad \quad \quad \tikzmark{revloopifstart}\mathsf{if} \; j = j' \; \mathsf{then} \\
    &\quad \quad \quad \quad \mathsf{yield}~y\tikzmark{revloopifend} \\
    &\quad \quad \mathsf{else} \\
    &\quad \quad \quad b \leftarrow \mathsf{true}\tikzmark{revloopexpend} \\
\end{align*}
\caption{Backward for loop expansion.}
\label{fig:revloopexp}
\end{wrapfigure}
\begin{equation*}
    s_\mathsf{rev} \ := \ \hlforRev{(j',y')}{ u }{\hlyield{y'}}
\end{equation*}
Some yield statements are reachable in
$s_\mathsf{rev}$, but not when iterating over $s$ in reverse order.
To ensure that we only output correct elements,
we replace every $\hlyield{\cdot}$ statement in $s_{\mathsf{rev}}$ by a
copy of $s$, leading to the programs $s' = s_{\mathsf{rev}}[ \hlyield{\cdot} \mapsto s ]$.
In our case:
\begin{align*}
    s' \ :=\ \hlforRev{(j',y')}{ u }{ s }
\end{align*}
It is now possible to replace every yield statement in this new program
by a conditional check ensuring that the output would actually be 
produced by the original program $s$.
\begin{equation*}
    s'' = s'[ \hlyield{e} \mapsto \hlifnoelse{\bcomp{i}{=}{j'}}{\hlyield{e}}]
\end{equation*}
In our case, the final program is described in \cref{fig:revloopexp}.
This rewriting can be generalised to any program of the form
$\hlforRev{(i,x)}{\ogen{s_1}}{s_2}$ combining the construction illustrated here
with the one taking care of position variables in the case of forward loops.
\section{Simple For Programs and Interpretations}
\label{sec:low-level}

In this section, we show how to compile a \kl{simple for-program} into
a \kl{first-order interpretation} in the symbolic setting. Recall that
this is already known to be theoretically possible in the non-symbolic case
\cite{bojanczyk2018polyregular}. However, this existing construction 
is not efficient: It requires computing a normal form of the \kl{simple for-program}
(\cite[Lemma~5.2]{bojanczyk2018polyregular}), and goes through
the model of pebble transducers \cite[Section~5]{bojanczyk2018polyregular} ---
both of these steps significantly increase the complexity of the generated formulas.

To transform a \kl{simple for-program} into a \kl{first-order interpretation},
we use as \kl{transduction tags} the set of all print statements in the program, remembering their
location in the source code. The \kl(formula){arity} of a print statement
is the number of the position variables present in its scope.
The \kl{output function} of a print statement is easy to define: if
the print statement outputs a fixed character $c$, then the \kl{output
function} returns $c$; otherwise,
if the print statement outputs $\mathsf{label}(i)$,
then the \kl{output function} returns the De Bruijn index \cite{DEBRUJ72}
of the variable $i$. For the \kl{ordering formula} between two print statements,
we use the technique for comparing addresses of the print statements, 
described in the \kl{for loop expansion} procedure:
In order to compare of two print statements, we compare 
their shared position variables,
breaking the ties using their ordering in the source code.
Observe that such \kl{ordering formulas} do not use quantifiers.

The hardest part is the \kl{domain formula}. This difficulty is akin to the one of the
\kl{for loop expansion} procedure for the reverse loop: given a print
statement $p(i_1, \dots, i_k)$, where $i_1, \dots, i_k$ are the position
variables in the scope of the print, we need to check whether it can
be reached. This amounts to taking the conjunction of the \texttt{if}-conditions, 
or their negations depending on the \texttt{if}-branch, along the path
from the root of the program to the print statement.
The only difficulty in defining this conjunction is
using the first-order logic to compute the values
of the boolean variables used in the \texttt{if}-conditions. 
We do this, by defining
\kl{program formulas}, which are \kl{first-order formulas} 
that describe how a program statement transforms
the values of its boolean variables.

\subsection{Program Formulas}
\label{sec:program-formulas}

\AP A \intro{program formula} is a \kl{first-order formula} where every
free variable is either: an \intro{input boolean variable}
$\ibvar{b}$, an \intro{output boolean variable} $\obvar{b}$, or an
\intro{input position variable} $\ipvar{i}$. In order to accommodate the
boolean variables, we introduce a new two-element sort $\mathbb{B}$. 
We handle it in the same way as the tag sort from \cref{sec:pullback}.

Given a fixed word $w \in \mathcal{D}^*$, a \kl{program formula}
$\varphi$ defines a relation between the input boolean variables
$\ibvar{b_1}, \dots, \ibvar{b_n}$, input position variables
$\ipvar{1}, \dots, \ipvar{k}$, and the output boolean variables
$\obvar{b_1}, \dots, \obvar{m}$. We are only interested in the \kl{program formulas}
that define \emph{functions} between the input and output variables, 
for every $w$.

In this section we show how to compute \kl{program formulas} for every
program statement $s$, that describes how the statement transforms its state.
The formulas are constructed inductively on the structure of the statement.
We start with the simplest case of \texttt{b := True}, whose program 
formula is defined as $\Phi_{\texttt{setTrue}} := \obvar{b}$.
Similarly, the program formula for a print statement is defined 
as $\Phi_{\texttt{print}} := \top$ (as it does not input or output any variables).
For the induction step, we need to consider three constructions: conditional
branching, sequencing, and iteration.

\paragraph{Conditional Branching.} 
Given two \kl{program formulas} $\Phi_1$ and $\Phi_2$
and a formula $\varphi$ that only uses input variables (position and booleans), 
we simulate the \texttt{if then else} construction in the following way:
\begin{equation*}
    \Phi_{\texttt{if}~\varphi~\texttt{then}~\Phi_1~\texttt{else}~\Phi_2} := (\varphi \land \Phi_1) \lor (\neg \varphi \land \Phi_2) \quad .
\end{equation*}
This construction only works if $\Phi_1$ and $\Phi_2$ have the same output variables. 
If this is not the case, we can to extend $\Phi_1$ and $\Phi_2$ with 
identity on the missing output variables, by adjoining them
with conjunctions of the form  $\ibvar{b} \iff \obvar{b}$
for each missing variable. 

\paragraph{Composition of Program Formulas.}
Let us consider two \kl{program formulas} $\Phi_1$ and $\Phi_2$,
and denote their input and output boolean variables as 
$B_1^{\mathsf{in}}, B_1^{\mathsf{out}}$ and $B_2^{\mathsf{in}}, B_2^{\mathsf{out}}$.
Let us start with the case where $B_2^{\mathsf{in}} = B_1^{\mathsf{out}} = \set{b_1, \dots, b_n}$.
In this case, we can compose the two program formulas in the following way:
\begin{equation*}
    \Phi_1 ; \Phi_2 :=
    \exists_{b_1 : \mathbb{B}} \cdots \exists_{b_n : \mathbb{B}}
    \quad
    \Phi_1[ \obvar{x} \mapsto x ]
    \wedge 
    \Phi_2[ \ibvar{x} \mapsto x ]
\end{equation*}
If the sets $B_1^{\mathsf{out}}$ and $B_2^{\mathsf{in}}$ are not equal,
we can deal with it by first ignoring every output variable $b$
of $\Phi_1$ that is not consumed by $\Phi_2$. Interestingly, this
requires an existential quantification:
$\Phi_1' \ := \ \exists_{b' : \mathbb{B}} \Phi_1[\obvar{b} \mapsto b']$.
Then, for each variable $b$ that is consumed by $\Phi_2$ but not produced by $\Phi_1$,
we add the identity clause ($\ibvar{b} \iff \obvar{b}$) to $\Phi_1'$ obtaining $\Phi_1''$.
After this modification, we can compose $\Phi_1''$ and $\Phi_2$ with no problems.

This definition of composition requires us to quantify over all variables form
$B_1^{\mathsf{out}} \cup B_2^{\mathsf{in}}$, which influences
the quantifier rank of the resulting program formula.
In our implementation, we are a bit more careful,
and only quantify over the variables from $B_1^{\mathsf{out}}
\cap (B_2^{\mathsf{in}} \cup B_2^{\mathsf{out}})$, obtaining the following bound:
\begin{equation*}
    \qrank(\Phi_1 ; \Phi_2) 
    \leq \max(\qrank(\Phi_1), \qrank(\Phi_2)) 
    +    |B_1^{\mathsf{out}} \cap (B_2^{\mathsf{in}} \cup B_2^{\mathsf{out}})|
    \quad .
\end{equation*}

\paragraph{Iteration of Program Formulas.} The most complex operation on 
program formulas is the iteration. We explain this on a representative 
case of a \kl{program formula} $\Phi$ which has a single \kl{input position variable} $\ipvar{i}$,
and whose \kl{output boolean variables} are the same as the \kl{input boolean variables}
($B^{\mathsf{in}} = B^{\mathsf{out}}$).

Given a word $w \in \mathcal{D}^*$, evaluating a forward
loop over $i$ in the range $0$ to $|w|$ amounts to the
following composition: 
\begin{equation}
    \label{eq:iteration-dumb}
    \Phi[\ipvar{i} \mapsto 0] ; \Phi[\ipvar{i} \mapsto 1] ; \cdots ;
    \Phi[\ipvar{i} \mapsto |w|] \quad ,
\end{equation}
The main difficulty is to compute this composition independently
of the length of the word $w$, while keeping the formula
and its \kl{quantifier rank} small.

To that end, we observe that $\Phi$ uses a 
finite number of boolean variables, and that each of those 
variables can only be set to \texttt{True} once (\ref{item:boolean-updates}).
As a consequence, in the composition in \cref{eq:iteration-dumb},
there are at most $|B^{\mathsf{out}}|$ steps that actually modify
the boolean variables. Based on this observation,
one can \emph{accelerate} the computation of the composition by
guessing the sequence of those steps ($p_1, \dots, p_{|B^{\mathsf{out}}|}$). 
The resulting \kl{program formula} $\Phi^*$ is given below
(we assume that $\Phi$ contains at least $3$ boolean variables,
and we denote their set as $\set{b_1, \dots, b_n}$ --
the cases for $n \leq 2$ are either analogous or trivial):
\begin{align}
    \Phi^* :=&~\exists_{ p_1 \leq \cdots \leq p_{n} : \mathbb{N}} 
    \label{eq:iteration-smart-pos}
    \\
             &~\exists_{\vec{b}_0, \vec{b}_1, \dots, \vec{b}_{n+1} : \mathbb{B}^n}
    \label{eq:iteration-smart-bool}
             \\
             &\bigwedge_{1 \leq j \leq n} \Phi(p_j ; \vec{b}_{j-1} ; \vec{b}_j)
    \label{eq:iteration-smart-correct}
             \\
             &\bigwedge_{1 \leq j \leq n+1}
               \forall_{p_{j-1} \leq p \leq p_j : \mathbb{N}} \;
               \Phi(p ; \vec{b}_{j-1} ; \vec{b}_{j-1}) 
    \label{eq:iteration-smart-complete}
    \\
             &\bigwedge_{1 \leq i \leq n} (\vec{b}_0)_i = \ibvar{b_i}
    \label{eq:iteration-smart-inital}
    \\
             &\bigwedge_{1 \leq i \leq n} (\vec{b}_{n+1})_i = \obvar{b_i} 
    \label{eq:iteration-smart-final}
            \quad .
\end{align}
The structure of this formula is as follows: In \cref{eq:iteration-smart-pos},
it guesses the steps $p_1, \dots, p_{n}$ that actually modify the boolean variables.
In \cref{eq:iteration-smart-bool}, it guesses the intermediate
values of the boolean variables ($\vec{b}_j$'s denote vectors of $n$ boolean variables). 
In \cref{eq:iteration-smart-correct}, it asserts that the guesses where \emph{correct}, 
i.e., that the \kl{program formula} $\Phi$ applied to position $p_j$ 
and the boolean variables $\vec{b}_{j-1}$ produces the boolean variables $\vec{b}_j$.
In \cref{eq:iteration-smart-complete}, it ensures that no position different than the $p_i$'s 
modifies the boolean variables.
(In this equation, $p_0$ and $p_{n+1}$ denote the first and the last position of the word.)
 Finally, in \cref{eq:iteration-smart-inital} and \cref{eq:iteration-smart-final},
it ensures that the initial and final values of the boolean variables are correctly set to the input and output values.
The formula for the reverse loop is similar, but guesses the positions $p_i$ in a decreasing order.

Our construction ensures the following bound on the quantifier rank of the resulting \kl{program formula}, 
which shows that the number of modified boolean variables is a crucial parameter for the complexity of the overall procedure:
\begin{equation}
    \label{eq:iteration-smart-quantifier-rank}
    \qrank(\Phi^*) 
    \leq \qrank(\Phi) 
    + |B^{\mathsf{out}}|^2
    + |B^{\mathsf{out}}|
    + 1 \quad .
\end{equation}
\section{Implementation}
\label{sec:benchmarks}

\AP We implemented all the transformations expressed in this paper in a
\texttt{Haskell} program. To measure the complexity of these transformations,
we associated to a \kl{high-level for-program} the following parameters: its
\emph{size} (number of control flow statements), its \intro{loop depth} (the
maximum number of nested loops), and its \intro{boolean depth} (the maximum
number of boolean variables visible at any point in the program). We compute
the same parameters for the corresponding \kl{simple for-program}. In the case
of \kl{first-order interpretations}, we only compute its \emph{size} (number of
nodes in the formula) and its \kl{quantifier rank}. This allowed us to
estimate the complexity of our transformations on a small set of programs that we present in
\cref{tab:benchmarks}. Then, we used several existing solvers to verify basic
\kl{first-order Hoare triples} for these programs. We
illustrate in \cref{tab:timings} the behaviour of the solvers on various
verification tasks, with a timeout of $5$ seconds for every solver.
These test offer only initial insight into the performance of our implementation,
so developing our implementation into an actual tool would require systematic 
benchmarks and comparison with already existing tools.

\begin{table}[t]
    \caption{Results for the transformations. 
        Here \kl[for-program]{FP} is a \kl{for-program},
        \kl[simple for-program]{S.FP} is a \kl{simple for-program},
        and \kl[first-order interpretation]{$\FO$-I} is a \kl{first-order interpretation}.
        The columns \textbf{l.d.}, \textbf{b.d.} and \textbf{q.r.}
        stand respectively for the \kl{loop depth}, 
        \kl{boolean depth} and \kl{quantifier rank}.
    }
    \label{tab:benchmarks}
    \centering
\setlength{\tabcolsep}{2mm}
\begin{tabular}{l|rrr|rrr|rrr}
    \toprule
     & \multicolumn{3}{c|}{\kl[for-program]{FP}} & \multicolumn{3}{c|}{\kl[simple for-program]{S.FP}} & \multicolumn{2}{c}{\kl[first-order interpretation]{$\FO$-I}} \\
    \textbf{filename} & \textbf{size} & \textbf{l.d.} & \textbf{b.d.} & \textbf{size} & \textbf{l.d.} & \textbf{b.d.} & \textbf{size} & \textbf{q.r.} \\
    \midrule
            \texttt{identity.pr} &
        $ 3 $ & $ 1 $ & $ 0 $ &
            $ 2    $ & $ 2  $ & $ 0 $  &
         $ 1 $ & $ 0 $  \\
            \texttt{reverse.pr} &
        $ 3 $ & $ 1 $ & $ 0 $ &
            $ 2    $ & $ 2  $ & $ 0 $  &
         $ 1 $ & $ 0 $  \\
            \texttt{subwords\_ab.pr} &
        $ 24 $ & $ 2 $ & $ 1 $ &
            $ 15    $ & $ 4  $ & $ 3 $  &
         $ 956 $ & $ 14 $  \\
            \texttt{map\_reverse.pr} &
        $ 36 $ & $ 2 $ & $ 1 $ &
            $ 18    $ & $ 4  $ & $ 1 $  &
         $ 285 $ & $ 5 $  \\
            \texttt{prefixes.pr} &
        $ 6 $ & $ 2 $ & $ 0 $ &
            $ 5    $ & $ 3  $ & $ 0 $  &
         $ 2 $ & $ 0 $  \\
            \texttt{get\_last\_word.pr} &
        $ 18 $ & $ 1 $ & $ 1 $ &
            $ 23    $ & $ 4  $ & $ 2 $  &
         $ 8553 $ & $ 15 $  \\
            \texttt{get\_first\_word.pr} &
        $ 22 $ & $ 1 $ & $ 1 $ &
            $ 5    $ & $ 2  $ & $ 0 $  &
         $ 103 $ & $ 4 $  \\
            \texttt{compress\_as.pr} &
        $ 12 $ & $ 1 $ & $ 1 $ &
            $ 12    $ & $ 3  $ & $ 2 $  &
         $ 209 $ & $ 10 $  \\
            \texttt{litteral\_test.pr} &
        $ 29 $ & $ 1 $ & $ 1 $ &
            $ 129    $ & $ 3  $ & $ 12 $  &
         $ 3.2 \times 10^4 $ & $ 82 $  \\
            \texttt{bibtex.pr} &
        $ 110 $ & $ 2 $ & $ 1 $ &
            $ 802    $ & $ 6  $ & $ 29 $  &
         $13.7 \times 10^6$ & $136$  \\
        \bottomrule
\end{tabular}
 \end{table}

\begin{table}
    \caption{Verification of \kl{first-order Hoare triples} over 
        sample \kl{for-programs}. We specify the preconditions and postconditions
        as regular languages, writing $\mathcal{L}_{ab}$ as a shorthand
        for $\mathcal{D}^*ab\mathcal{D}^*$,
        and similarly for $\mathcal{L}_{aa}$, $\mathcal{L}_{ba}$, etc.
        In the columns corresponding to the solvers, a checkmark
        indicates a positive reply,
        a cross mark indicates a negative reply,
        and a question mark indicates a timeout
        or a memory exhaustion.
        We indicate the size and the \kl{quantifier rank} (\textbf{q.r.})
        of the \kl{first-order formulas} that are fed to the solvers.
    }
    \label{tab:timings}
    \centering
\setlength{\tabcolsep}{2mm}
\begin{tabular}{l|l|l|r|r|r|r|r}
    \toprule
\textbf{Name} & \textbf{Pre.} & \textbf{Post.} & \textbf{q.r.} & \textbf{size} & \textbf{MONA} &  \textbf{CVC5} & \textbf{Z3} \\
\midrule
\texttt{compress\_as.pr} & $\mathcal{L}_{ab}$ & $\mathcal{L}_{ab}$ & 16 & 763 & \OK & 
\UK & \UK \\
\texttt{reverse\_add\_hash.pr} & $\mathcal{L}_{ab}$ & $\mathcal{L}_{ba}$ & 9 & 380 & \UK & 
\OK & \UK \\
\texttt{get\_last\_word.pr} & $\mathcal{D}^*a$ & $\mathcal{L}_{aa}$ & 27 & 28274 & \UK & 
\UK & \KO \\
\texttt{subwords\_ab.pr} & $\mathcal{L}_{ab}$ & $\mathcal{L}_{ab}$ & 26 & 3276 & \textcolor{C3}{\UK} & 
\UK & \UK \\
\texttt{map\_reverse.pr} & $\mathcal{D}^*a$ & $a \mathcal{D}^*$& 13 & 801 & \UK & 
\UK & \UK \\
    \bottomrule
\end{tabular}
 \end{table}

\paragraph{Compilation to $\FO$-formulas.}
Looking at \cref{tab:benchmarks}, we observe that the generated simple for programs 
have reasonable size and boolean depth. The generated \kl{first-order interpretations}
still have reasonable quantifier ranks, but their size grows significantly.
In the simplest cases of \cref{tab:benchmarks}, our compilation procedure is able
to eliminate all boolean variables, thus producing a \emph{quantifier-free}
formula. This is the case for \texttt{identity.pr}, \texttt{reverse.pr} and
\texttt{prefixes.pr}. Moreover, we observe that the \kl{boolean depth} of the
\kl{simple for-program} is a good indicator of the \kl{quantifier rank}
of the generated \kl{first-order interpretation}. Furthermore, the tests indicate that
\kl{elimination of literals} is responsible for a significant increase of the formulas 
size and quantifier rank (\texttt{literal\_test.pr} and \texttt{bibtex.pr}).  
This is explained by the fact that the elimination of literals introduces
(non-cyclic) counters, simulated by a number of boolean variables. Finally,
we observe that the size of the generated formulas differs significantly
for the programs \texttt{get\_first\_word.pr} and \texttt{get\_last\_word.pr}.
This is somewhat surprising, as the two programs are symmetric with respect to 
reversing the input words, and indicates some room for improvement in handling 
the reversed iteration.

\paragraph{Solver Performance.} We can observe in \cref{tab:timings} that the
different solvers are complementary. This might seem surprising, as the
\kl{MONA} solver is a complete decision procedure. However, since it solves a
problem that is \TOWER-complete \cite[Theorem 13.5]{REINH02}, it is
understandable that it underperforms the SMT solvers on some instances, even
though we use them with the undecidable \texttt{UFDTLIA} theory. Let us justify
this choice of the SMTLib theory: (a) \emph{Uninterpreted Functions} (\texttt{UF}) are
used to represent the word, which is treated as a function from positions to
characters, (b) \emph{Data Types} (\texttt{DT}) is used to represent finite sets of tags
and characters, (c) \emph{Linear Integer Arithmetic} (\texttt{LIA}) is used to deal with
the order of the positions in the word. This choice might not be optimal, but
we believe that it is a good trade off between ease-of-use and performance for
our proof-of-concept implementation. We can also observe that no solvers was
able to deal with \texttt{subwords\_ab.pr} and \texttt{map\_reverse.pr} within
the $5$ seconds timeout. Understanding the complexities that arise in those
cases, might be helpful for improving the performance of our implementation.
\section{Conclusion}
\label{sec:conclusion}

We have show that the theory of \kl{star-free polyregular functions} can be
used to verify close to real-world programs, and have implemented a prototype
tool that can discharge simple verification goals to existing solvers.

\paragraph{Benchmarks.} It would be interesting to systematically benchmark our
implementation against existing tools for verifying \kl{linear regular
functions}, in the case of first order specifications. Since our approach
allows for polynomial size transformations, it would also be interesting to
devise a set of benchmarks for this broader classe of functions.

\paragraph{Optimizations.} The preliminary tests indicate that one of the most
promising source of optimizations is managing the \kl{boolean depth} of the
generated \kl{simple for-programs} during compilation. This can be achieved by
post-compilation optimizations (constant propagation, dead code elimination),
or by improving the code generation mechanism itself, which are low-hanging
fruits for future work. One source of the boolean variables seems to be the
\kl{elimination of Literal Equality} step (\ref{item:lit_elim}), which could be
mitigated by adding explicit successor and predecessor predicates to the
language of \kl{simple for-programs}.

At the level of \kl{first-order interpretations}, we have identified several
directions for improving their efficiency. One optimization is computing the
sequential composition of programs in a way that minimizes the number of
quantified boolean variables. Similarly, there seems to be potential for
performing direct substitutions instead of quantifying over the variables in a
lot of cases. Finally, our current approach for handling loops introduces
universal quantifiers, whose number could be reduced by exploiting the
monotonicity of the state transformations.
    
\paragraph{Solver Integration.} There is a lot of potential for optimizing the
input and parameters of the solvers for our particular use-case. An interesting
research direction would be to reduce the verification problem to emptiness of
LTL formulas, allowing us to use LTL solvers such as \texttt{SPOT} \cite{SPOT}.

\paragraph{Modular Verification.} The benchmarks show that one of the main
bottlenecks of our approach is the expansion of loops (whether in the
translation to \kl{simple for-programs} or in the translation to
\kl{first-order interpretations}). For this reason, the ability to verify
statements of the form \texttt{for (i, e) in enumerate(f(x)) do s done}, based
on a specification of $f$ given as a Hoare triple, would be a significant
improvement. However, it remains unclear how to integrate such modular
verification in our current approach.

\paragraph{Language Design.} As mentioned in
\cref{sec:high-level}, \kl{for-programs}
extended with unrestricted booleans also enjoy a decidable verification of
Hoare triples. However, the verification algorithm uses of monadic second-order
logic (MSO) over words instead of first-order logic. While this prohibits the
use of traditional SMT solvers, this logic can be handled by the \kl{MONA}
solver, and it might be interesting to implement and test the unrestricted
version of the language. 

Another interesting extension of the language would be to allow the use of
complex types, such as pairs and records. This would make the language closer
to real use cases such as configuration management and data processing. It
would require extending the specification language to structured data types,
bypassing the current limitation that we can only verify string-to-string
transformations.

\paragraph{Integration with Existing Tools.} It would be a natural next step to
integrate our results inside frameworks for program verification or testing.
This could be by checking goals generated by a tool such as \texttt{Why3}
\cite{Why3}, or by verifying properties of Python programs using decorated
functions. We would also like to point out that verification methods based on a
\kl{regularity preserving property} (such as done in \cite{CHLRW19}) can
transparently use our more general class of programs as input, instead of the
more traditional \kl{linear regular functions}. 

\section*{Acknowledgements}
We would like to thank Arnav Garg and Ojas Maheshwari for their participation in the early stage of this project.

\bibliographystyle{plainurl}
\appendix
\section{Appendix}

\begin{lemma}
    \label{lem:umc-equality-nested-words}
    Allowing unrestricted equality checks between two \kl{nested words}
    results in the undecidability of the model checking problem.
\end{lemma}
\begin{proof}
    For every instance of the Post Correspondence Problem (PCP), we can 
    construct a function \texttt{f(x : list[list[Char]]) : Bool}
    in the \kl{high-level language} with unrestricted equality checks, such 
    that $f(x) = \btrue$ if and only if $x$ encodes a solution to the PCP instance.
    For example the PCP instance $\{ (ab, a), (b, aa), (ba, b) \}$ can be encoded 
    as the following function:
\begin{Shaded}
\begin{Highlighting}[]
\KeywordTok{def}\NormalTok{ top(x : }\BuiltInTok{list}\NormalTok{[}\BuiltInTok{list}\NormalTok{[}\BuiltInTok{chr}\NormalTok{]]) }\OperatorTok{{-}\textgreater{}} \BuiltInTok{list}\NormalTok{[}\BuiltInTok{chr}\NormalTok{]:}
    \ControlFlowTok{for}\NormalTok{ elem }\KeywordTok{in}\NormalTok{ x: }
        \ControlFlowTok{if}\NormalTok{ elem }\OperatorTok{==} \StringTok{"one"}\NormalTok{:}
            \ControlFlowTok{yield} \StringTok{"a"}
            \ControlFlowTok{yield} \StringTok{"b"}
        \ControlFlowTok{elif}\NormalTok{ elem }\OperatorTok{==} \StringTok{"two"}\NormalTok{:}
            \ControlFlowTok{yield} \StringTok{"a"}
        \ControlFlowTok{elif}\NormalTok{ elem }\OperatorTok{==} \StringTok{"three"}\NormalTok{:}
            \ControlFlowTok{yield} \StringTok{"b"}
            \ControlFlowTok{yield} \StringTok{"a"}

\KeywordTok{def}\NormalTok{ bottom(x : }\BuiltInTok{list}\NormalTok{[}\BuiltInTok{list}\NormalTok{[}\BuiltInTok{chr}\NormalTok{]]) }\OperatorTok{{-}\textgreater{}} \BuiltInTok{list}\NormalTok{[}\BuiltInTok{chr}\NormalTok{]:}
    \ControlFlowTok{for}\NormalTok{ elem }\KeywordTok{in}\NormalTok{ x: }
        \ControlFlowTok{if}\NormalTok{ elem }\OperatorTok{==} \StringTok{"one"}\NormalTok{: }
            \ControlFlowTok{yield} \StringTok{"a"}
        \ControlFlowTok{elif}\NormalTok{ elem }\OperatorTok{==} \StringTok{"two"}\NormalTok{:}
            \ControlFlowTok{yield} \StringTok{"a"}
            \ControlFlowTok{yield} \StringTok{"a"}
        \ControlFlowTok{elif}\NormalTok{ elem }\OperatorTok{==} \StringTok{"three"}\NormalTok{:}
            \ControlFlowTok{yield} \StringTok{"b"}

\KeywordTok{def}\NormalTok{ pcp(x : }\BuiltInTok{list}\NormalTok{[}\BuiltInTok{list}\NormalTok{[}\BuiltInTok{str}\NormalTok{]]) }\OperatorTok{{-}\textgreater{}} \BuiltInTok{bool}\NormalTok{ :}
    \ControlFlowTok{return} \BuiltInTok{list}\NormalTok{(top(x)) }\OperatorTok{==} \BuiltInTok{list}\NormalTok{(bottom(x))}
\end{Highlighting}
\end{Shaded}
\end{proof}

\begin{lemma}
    \label{lem:fo-emptiness}
    \proofref{lem:fo-emptiness}
    The \kl{emptiness} problem for the \kl{first-order logic on words} is decidable for the infinite alphabet $\mathcal{D}$.
\end{lemma}
\begin{proof}
    Take a formula $\varphi$ and observe that is contains only a finite number of constants from $\mathcal{D}$ -- call this set $A$.
    It is not hard to see that the truth value of $\varphi$ is \kl{supported} by $A$: for every function $f : \mathcal{D} \to \mathcal{D}$
    that does not touch 
    elements of $A$, the truth value of $\varphi$ is the same for $w$ and $f^*(w)$. (Remember that 
    $f^*$ is the pointwise application of $f$). 
    Let $\mathtt{blank} \in \mathcal{D}$ be a letter that does not appear in $A$,
    and observe that the formula $\varphi$ is satisfied for some word in $\mathcal{D}^*$ if and only if it is satisfied by
    some word in $(A \cup \{\mathtt{blank}\})^*$. Indeed, if we take a function $g: \mathcal{D} \to \mathcal{D}$ that does not touch elements of $A$
    and maps all other letters to $\mathtt{blank}$, we can use it to map $\mathcal{D}^*$ to $(A \cup \{\mathtt{blank}\})^*$ in a way 
    that preserves the truth value of $\varphi$.
    This finishes the proof of the lemma, as we have reduced the general problem to a finite alphabet $A \cup \{\mathtt{blank}\}$.
\end{proof}

\begin{figure}[h]
    \centering
\begin{Shaded}
\begin{Highlighting}[numbers=left]
\KeywordTok{def}\NormalTok{ getBetween( l : }\DataTypeTok{[Char]} \KeywordTok{with}\NormalTok{ (i,j) ) : }\DataTypeTok{[Char]}\NormalTok{ := }
\KeywordTok{    for}\NormalTok{ (k,c) }\KeywordTok{in} \KeywordTok{enumerate}\NormalTok{(l) }\KeywordTok{do}
        \KeywordTok{if}\NormalTok{ i \textless{}= k }\KeywordTok{and}\NormalTok{ k \textless{}= j }\KeywordTok{then}
            \KeywordTok{yield}\NormalTok{ c}
        \KeywordTok{endif}
    \KeywordTok{done}

\KeywordTok{def}\NormalTok{ containsAB(w : }\DataTypeTok{[Char]}\NormalTok{) : }\DataTypeTok{Bool}\NormalTok{ := }
    \KeywordTok{let} \KeywordTok{mut}\NormalTok{ seen\_a := }\KeywordTok{False} \KeywordTok{in} 
\KeywordTok{    for}\NormalTok{ (i, c) }\KeywordTok{in} \KeywordTok{enumerate}\NormalTok{(w) }\KeywordTok{do}
    \KeywordTok{if}\NormalTok{ c === }\StringTok{\textquotesingle{}a\textquotesingle{} }\KeywordTok{then}
\NormalTok{            seen\_a := }\KeywordTok{True}
    \KeywordTok{else} \KeywordTok{if}\NormalTok{ c === }\StringTok{\textquotesingle{}b\textquotesingle{} }\KeywordTok{and}\NormalTok{ seen\_a }\KeywordTok{then}
            \KeywordTok{return} \KeywordTok{True}
        \KeywordTok{endif} \KeywordTok{endif}
    \KeywordTok{done}
    \KeywordTok{return} \KeywordTok{False}

\KeywordTok{def}\NormalTok{ subwordsWithAB(w : }\DataTypeTok{[Char]}\NormalTok{) : }\DataTypeTok{[[Char]]}\NormalTok{ := }
\KeywordTok{    for}\NormalTok{ (i,c) }\KeywordTok{in} \KeywordTok{enumerate}\NormalTok{(w) }\KeywordTok{do}
\KeywordTok{        for}\NormalTok{ (j,d) }\KeywordTok{in} \KeywordTok{reversed}\NormalTok{(}\KeywordTok{enumerate}\NormalTok{(w)) }\KeywordTok{do}
            \KeywordTok{let}\NormalTok{ s := getBetween(w }\KeywordTok{with}\NormalTok{ (i,j)) }\KeywordTok{in}
            \KeywordTok{if}\NormalTok{ containsAB(s) }\KeywordTok{then}
                \KeywordTok{yield}\NormalTok{ s}
            \KeywordTok{endif}
        \KeywordTok{done}
    \KeywordTok{done}

\KeywordTok{def}\NormalTok{ main (w : }\DataTypeTok{[Char]}\NormalTok{) : }\DataTypeTok{[Char]}\NormalTok{ := }
    \KeywordTok{let}\NormalTok{ subwrds := subwordsWithAB(w) }\KeywordTok{in}
\KeywordTok{    for}\NormalTok{ (j,s) }\KeywordTok{in} \KeywordTok{enumerate}\NormalTok{(subwrds) }\KeywordTok{do}
\KeywordTok{        for}\NormalTok{ (i,c) }\KeywordTok{in} \KeywordTok{enumerate}\NormalTok{(s) }\KeywordTok{do}
            \KeywordTok{yield}\NormalTok{ c}
        \KeywordTok{done}
        \KeywordTok{yield}\StringTok{ \textquotesingle{}\#\textquotesingle{}}
    \KeywordTok{done}
\end{Highlighting}
\end{Shaded}
\caption{The \kl{for-program} computing all subwords of a word containing the substring $ab$,
corresponding to the Python code in \cref{fig:python-example-nested}.}
\label{fig:high-level-example-nested}
\end{figure}

\begin{figure}[h]
    \centering
\begin{Shaded}
\begin{Highlighting}[numbers=left]
\KeywordTok{for}\NormalTok{ i }\KeywordTok{in} \KeywordTok{input} \KeywordTok{do}
    \KeywordTok{for}\NormalTok{ j }\KeywordTok{in} \KeywordTok{reversed}\NormalTok{(}\KeywordTok{input}\NormalTok{) }\KeywordTok{do}
        \KeywordTok{let}\NormalTok{ b2, b3, b4 := }\KeywordTok{false} \KeywordTok{in}
        \KeywordTok{for}\NormalTok{ k }\KeywordTok{in} \KeywordTok{input} \KeywordTok{do}
            \KeywordTok{if}\NormalTok{ (i <= k) and (k <= j) }\KeywordTok{then}
                \KeywordTok{if}\NormalTok{ label(k) == \textquotesingle{}a\textquotesingle{} }\KeywordTok{then}
\NormalTok{                    b4 := }\KeywordTok{true}
                \KeywordTok{else}
                    \KeywordTok{if}\NormalTok{ (label(k) == \textquotesingle{}b\textquotesingle{}) and (b4) }\KeywordTok{then}
                        \KeywordTok{if}\NormalTok{ b3 }\KeywordTok{then}
                            \KeywordTok{skip}
                        \KeywordTok{else}
\NormalTok{                            b3 := }\KeywordTok{true}
\NormalTok{                            b2 := }\KeywordTok{true}
                        \KeywordTok{endif}
                    \KeywordTok{else}
                        \KeywordTok{skip}
                    \KeywordTok{endif}
                \KeywordTok{endif}
            \KeywordTok{else}
                \KeywordTok{skip}
            \KeywordTok{endif}
        \KeywordTok{done}
        \KeywordTok{if}\NormalTok{ b2 }\KeywordTok{then}
            \KeywordTok{for}\NormalTok{ l }\KeywordTok{in} \KeywordTok{input} \KeywordTok{do}
                \KeywordTok{if}\NormalTok{ (i <= l) and (l <= j) }\KeywordTok{then}
                    \KeywordTok{print}\NormalTok{ label(l)}
                \KeywordTok{else}
                    \KeywordTok{skip}
                \KeywordTok{endif}
            \KeywordTok{done}
            \KeywordTok{print}\NormalTok{ \textquotesingle{}\#\textquotesingle{}}
        \KeywordTok{else}
            \KeywordTok{skip}
        \KeywordTok{endif}
     \KeywordTok{done}
\KeywordTok{done}
\end{Highlighting}
\end{Shaded}
\caption{The \kl{simple for-program} computing all subwords of a word containing the substring $ab$,
corresponding to the Python code in \cref{fig:python-example-nested}, and obtained
by compiling \cref{fig:high-level-example-nested}.}
\label{fig:low-level-example-nested}
\end{figure}

\begin{figure}[h]
    \centering
\begin{Shaded}
\begin{Highlighting}[]
\KeywordTok{def}\NormalTok{ eq(u, v):}
    \ControlFlowTok{for}\NormalTok{ (i, ui) }\KeywordTok{in} \BuiltInTok{enumerate}\NormalTok{(u):}
        \ControlFlowTok{for}\NormalTok{ (j, vj) }\KeywordTok{in} \BuiltInTok{enumerate}\NormalTok{(v):}
            \ControlFlowTok{if}\NormalTok{ i }\OperatorTok{==}\NormalTok{ j }\KeywordTok{and}\NormalTok{ ui }\OperatorTok{!=}\NormalTok{ vj:}
                \ControlFlowTok{return} \VariableTok{False}
    \ControlFlowTok{return} \VariableTok{True}
\end{Highlighting}
\end{Shaded}
\caption{Encoding the equality of two words $u$ and $v$ in Python,
using a comparison between indices of two different lists.}
\label{fig:eq-def-different-indices}
\end{figure}

\begin{figure}[h]
    \centering
\begin{Shaded}
\begin{Highlighting}[]
\KeywordTok{def}\NormalTok{ switch(b, u, v):}
    \ControlFlowTok{if}\NormalTok{ b:}
        \ControlFlowTok{return}\NormalTok{ u}
    \ControlFlowTok{else}\NormalTok{:}
        \ControlFlowTok{return}\NormalTok{ v}

\KeywordTok{def}\NormalTok{ eq(u, v):}
\NormalTok{    b }\OperatorTok{=} \VariableTok{False}
    \ControlFlowTok{for}\NormalTok{ (i, ui) }\KeywordTok{in} \BuiltInTok{enumerate}\NormalTok{(switch(b, u, v)):}
\NormalTok{        b }\OperatorTok{=} \VariableTok{True}
        \ControlFlowTok{for}\NormalTok{ (j, vj) }\KeywordTok{in} \BuiltInTok{enumerate}\NormalTok{(switch(b, u, v)):}
            \ControlFlowTok{if}\NormalTok{ i }\OperatorTok{==}\NormalTok{ j }\KeywordTok{and}\NormalTok{ ui }\OperatorTok{!=}\NormalTok{ vj:}
                \ControlFlowTok{return} \VariableTok{False}
    \ControlFlowTok{return} \VariableTok{True}
\end{Highlighting}
\end{Shaded}
\caption{Encoding the equality of two words $u$ and $v$ in Python,
using a function taking a boolean as input.}
\label{fig:eq-def-boolean}
\end{figure}

\begin{figure}[h]
    \centering
\begin{Shaded}
\begin{Highlighting}[]
\KeywordTok{def}\NormalTok{ eq(u, v):}
\NormalTok{    w }\OperatorTok{=}\NormalTok{ u}
    \ControlFlowTok{for}\NormalTok{ (i, ui) }\KeywordTok{in} \BuiltInTok{enumerate}\NormalTok{(w):}
\NormalTok{        w }\OperatorTok{=}\NormalTok{ v}
        \ControlFlowTok{for}\NormalTok{ (j, vj) }\KeywordTok{in} \BuiltInTok{enumerate}\NormalTok{(w):}
            \ControlFlowTok{if}\NormalTok{ i }\OperatorTok{==}\NormalTok{ j }\KeywordTok{and}\NormalTok{ ui }\OperatorTok{!=}\NormalTok{ vj:}
                \ControlFlowTok{return} \VariableTok{False}
    \ControlFlowTok{return} \VariableTok{True}
\end{Highlighting}
\end{Shaded}
\caption{Encoding the equality of two words $u$ and $v$ in Python,
using the shadowing of a variable to switch between two lists.}
\label{fig:eq-def-shadowing}
\end{figure}

\begin{figure}[h]
    \centering
    \begin{align*}
        \intro*\bBinOp := &~ \land ~|~ \lor ~|~ \Rightarrow ~|~ \Leftrightarrow \\
        \intro*\pCmpOp := &~ = ~|~ \neq ~|~ < ~|~ \leq ~|~ > ~|~ \geq \\
        \intro*\bexpr :=&~ \intro*\btrue ~|~ \intro*\bfalse ~|~ \intro*\bnot{\bexpr} \\
               |&~ \bbin{\bexpr}{\bBinOp}{\bexpr}   \\
               |&~ \bcomp{i}{\pCmpOp}{j} & i,j \in \PVars \\
               |&~ f(\bexpr) & f \in \FunVars \\
               |&~ \intro*\bliteq{\cexpr}{\oexpr}
    \end{align*}
    \caption{The syntax of \kl{boolean expressions}.}
    \label{fig:bool-expr}
\end{figure}

\begin{figure}[h]
    \centering
    \begin{align*}
        \intro*\cexpr :=&~ \mathsf{char} \; c & c \in \Sigma \\
               |&~ \mathsf{list}(\cexpr_1, \ldots, \cexpr_n)
    \end{align*}
    \caption{The syntax of \kl{constant expressions}.}
    \label{fig:const-expr}
\end{figure}

\begin{figure}[h]
    \centering
    \begin{align*}
        \intro*\oexpr :=&~ x & x \in \OVars \\
               |&~ \cexpr \\
               |&~ \intro*\olist{\oexpr_1, \dots,  \oexpr_n}  \\
               |&~ f(\aexpr_1, \dots, \aexpr_n) & f \in \FunVars \\
    \end{align*}
    \caption{The syntax of \kl{list expressions}.}
    \label{fig:out-expr}
\end{figure}

\begin{figure}[h]
    \centering
    \AP
    \begin{align*}
        \intro*\hlstmt :=&~ 
                   \intro*\hlif{\bexpr}{\hlstmt}{\hlstmt} \\
               |&~ \intro*\hlyield{\oexpr} \\
               |&~ \intro*\hlreturn{\oexpr} \\
               |&~ \intro*\hlletoutput{x}{\oexpr}{\hlstmt} & x \in \OVars \\
               |&~ \intro*\hlletboolean{x}{\hlstmt} & x \in \BVars \\
               |&~ \intro*\hlsettrue{x} & x \in \BVars \\
               |&~ \intro*\hlfor{(i,x)}{\oexpr}{\hlstmt} & (i,x) \in \PVars \times \OVars \\
               |&~ \intro*\hlforRev{(i,x)}{\oexpr}{\hlstmt} & (i,x) \in \PVars \times \OVars \\
               |&~ \intro*\hlseq{\hlstmt}{\hlstmt}
    \end{align*}
    \caption{The syntax of \kl{high-level control statements}.}
    \label{fig:high-level-stmt}
\end{figure}

\begin{figure}[h]
    \centering
    \begin{align*}
        \intro*\aexpr :=&~ (\oexpr, p_1, \dots, p_n) & \forall 1 \leq i \leq n, p_i \in \PVars \\
        \intro*\hlfun :=&~ \hlfundef{f}{\aexpr_1, \dots, \aexpr_n}{\hlstmt} & f \in \FunVars \\
        \intro*\hlprogram :=&~ ([\hlfun_1, \dots, \hlfun_n], f) & f \in \FunVars \\
    \end{align*}
    \caption{The syntax of \kl{high-level for-programs}.}
    \label{fig:high-level-program}
\end{figure}

\begin{figure}[h]
\begin{align*}
    \mathsf{arg} ::=&~ (\TOut[n],\ell) & \ell \in \Nat \\
    \mathsf{fun} ::=&~ 
           \mathsf{arg}_1 \times \cdots \times \mathsf{arg}_k \to \TBool \\
    \mid&~ \mathsf{arg}_1 \times \cdots \times \mathsf{arg}_k \to \TOut[n] 
\end{align*}
\caption{Possible types of \kl{for-programs} and their functions.}
\label{fig:typing-for-programs}
\end{figure}

\begin{figure}[h]
    \begin{prooftree}
    \AxiomC{}
    \RightLabel{(T-True)}
    \UnaryInfC{$\Gamma \vdash \btrue : \TBool$}
    \end{prooftree}

    \begin{prooftree}
    \AxiomC{}
    \RightLabel{(T-False)}
    \UnaryInfC{$\Gamma \vdash \bfalse : \TBool$}
    \end{prooftree}

    \begin{prooftree}
    \AxiomC{$\Gamma \vdash e : \TBool$}
    \RightLabel{(T-Not)}
    \UnaryInfC{$\Gamma \vdash \bnot{e} : \TBool$}
    \end{prooftree}

    \begin{prooftree}
    \AxiomC{$\Gamma \vdash e_1 : \TBool$}
    \AxiomC{$\Gamma \vdash e_2 : \TBool$}
    \RightLabel{(T-BBin)}
    \BinaryInfC{$\Gamma \vdash \bbin{e_1}{op}{e_2} : \TBool$}
    \end{prooftree}

    \begin{prooftree}
    \AxiomC{$\Gamma \vdash i :  \TPos[o_i]$}
    \AxiomC{$\Gamma \vdash j : \TPos[o_j]$}
    \AxiomC{$o_i = o_j$}
    \RightLabel{(T-PComp)}
    \TrinaryInfC{$\Gamma \vdash \bcomp{i}{op}{j} : \TBool$}
    \end{prooftree}

    \caption{Typing rules for boolean expressions.}
    \label{fig:typing-bool}
\end{figure}

\begin{figure}[h]
    \begin{prooftree}
    \AxiomC{}
    \RightLabel{(T-OVar)}
    \UnaryInfC{$\Gamma, x : \TOut[n] \vdash x : \TOut[n]$}
    \end{prooftree}

    \begin{prooftree}
        \AxiomC{$\Gamma \vdash e_i : \TOut[n]$ for all $i$}
    \RightLabel{(T-OList)}
    \UnaryInfC{$\Gamma \vdash \olist{e_1,\ldots,e_n} : \TOut[n+1]$}
    \end{prooftree}

    \caption{Typing rules for \kl{list expressions} and \kl{constant expressions}.}
    \label{fig:typing-output}
\end{figure}

\begin{figure}[h]
\begin{prooftree}
\AxiomC{$\Gamma \vdash e : \TBool$}
\AxiomC{$\Gamma \vdash s_1 : \tau$}
\AxiomC{$\Gamma \vdash s_2 : \tau$}
\RightLabel{(T-If)}
\TrinaryInfC{$\Gamma \vdash \hlif{e}{s_1}{s_2} : \tau$}
\end{prooftree}

\begin{prooftree}
\AxiomC{$\Gamma \vdash e : \TOut[n]$}
\RightLabel{(T-Yield)}
\UnaryInfC{$\Gamma \vdash \hlyield{e} : \TOut[n+1]$}
\end{prooftree}

\begin{prooftree}
\AxiomC{$\Gamma \vdash e : \TOut[n]$}
\RightLabel{(T-Return)}
\UnaryInfC{$\Gamma \vdash \hlreturn{e} : \TOut[n]$}
\end{prooftree}

\begin{prooftree}
\AxiomC{$\Gamma \vdash e : \TBool$}
\RightLabel{(T-Return)}
\UnaryInfC{$\Gamma \vdash \hlreturn{e} : \TBool$}
\end{prooftree}

\begin{prooftree}
\AxiomC{$\Gamma \vdash e : \TOut[n]$}
\AxiomC{$\Gamma, x:\TOut[n] \vdash s : \tau$}
\AxiomC{$x \not\in \Gamma$}
\RightLabel{(T-LetOut)}
\TrinaryInfC{$\Gamma \vdash \hlletoutput{x}{e}{s} : \tau$}
\end{prooftree}

\begin{prooftree}
\AxiomC{$\Gamma, x:\TBool \vdash s : \tau$}
\AxiomC{$x \not\in \Gamma$}
\RightLabel{(T-LetBool)}
\BinaryInfC{$\Gamma \vdash \hlletboolean{x}{s} : \tau$}
\end{prooftree}

\begin{prooftree}
\AxiomC{$\Gamma \vdash o : \TOut[n]$}
\AxiomC{$\Gamma, i:\TPos[o], x:\TOut[n] \vdash s : \tau$}
\AxiomC{$n > 0$}
\RightLabel{(T-For)}
\TrinaryInfC{$\Gamma \vdash \hlfor{(i,x)}{o}{s} : \tau$}
\end{prooftree}

\begin{prooftree}
\AxiomC{$\Gamma \vdash o : \TOut[n]$}
\AxiomC{$\Gamma, i:\TPos[o], x:\TOut \vdash s : \tau$}
\AxiomC{$n > 0$}
\RightLabel{(T-ForRev)}
\TrinaryInfC{$\Gamma \vdash \hlforRev{(i,x)}{o}{s} : \tau$}
\end{prooftree}

\begin{prooftree}
\AxiomC{}
\RightLabel{(T-SetTrue)}
\UnaryInfC{$\Gamma \vdash \hlsettrue{x} : \tau$}
\end{prooftree}

\begin{prooftree}
\AxiomC{$\Gamma \vdash s_1 : \tau$}
\AxiomC{$\Gamma \vdash s_2 : \tau$}
\RightLabel{(T-Seq)}
\BinaryInfC{$\Gamma \vdash \hlseq{s_1}{s_2} : \tau$}
\end{prooftree}
\caption{Typing rules for control statements.}
\label{fig:typing-control}
\end{figure}

\begin{figure}[h]
\begin{prooftree}
    \AxiomC{$\Gamma \vdash o : \TOut[n]$}
    \AxiomC{$\Gamma \vdash p_i : \TPos[o]$ for all $1 \leq i \leq n$}
    \RightLabel{(T-AExpr)}
    \BinaryInfC{$\Gamma \vdash (o, p_1, \dots, p_n) : (\TOut[n], n)$}
\end{prooftree}

\begin{prooftree}
\AxiomC{$\Gamma \vdash a_i : \tau_i$ for all $i$}
\AxiomC{$\Gamma \vdash s : \tau$}
\RightLabel{(T-Fun)}
\BinaryInfC{$\Gamma \vdash \hlfundef{f}{a_1, \dots, a_n}{s} : (\tau_1,\ldots,\tau_n) \to \tau$}
\end{prooftree}

\begin{prooftree}
    \AxiomC{$\Gamma \vdash f : (\tau_1,\ldots,\tau_n) \to \tau$}
    \AxiomC{$\Gamma \vdash a_i : \tau_i$ for all $1 \leq i \leq n$}
    \RightLabel{(T-App)}
    \BinaryInfC{$\Gamma \vdash f(a_1,\ldots,a_n) : \tau$}
\end{prooftree}

\begin{prooftree}
\AxiomC{$\Gamma, \seqof{f_j : \tau_j}[j < i] \vdash f_i : \tau_i$ for all $1 \leq i \leq n$}
\AxiomC{$f = f_j$ for some $1 \leq j \leq n$}
\RightLabel{(T-Prog)}
\BinaryInfC{$\Gamma \vdash ([f_1, \dots, f_n], f) : \tau_j$}
\end{prooftree}
\caption{Typing rules of \kl{for-programs}.}
\label{fig:typing-high-level}
\end{figure}

\begin{figure}[h]
    \centering

\begin{prooftree}
    \AxiomC{$\Gamma' \vdash s : \TBool$}
    \AxiomC{$\Gamma' \subseteq \Gamma$}
    \AxiomC{$\Gamma'$ contains no boolean variables}
    \RightLabel{(B-Gen)}
    \TrinaryInfC{$\Gamma \vdash \bgen{ s } \colon \TBool$}
\end{prooftree}

\begin{prooftree}
    \AxiomC{$\Gamma' \vdash s : \TOut[n]$}
    \AxiomC{$\Gamma' \subseteq \Gamma$}
    \AxiomC{$\Gamma'$ contains no boolean variables}
    \RightLabel{(L-Gen)}
    \TrinaryInfC{$\Gamma \vdash \ogen{ s } \colon \TOut[n]$}
\end{prooftree}

    \caption{The syntax of \kl{generator expressions}.}
    \label{fig:generators}
\end{figure}

\end{document}